\documentclass{aa}
\usepackage{graphicx}
\begin{document}
%
%
\title{Using Cepheids to determine the Galactic abundance gradient. I.
The solar neighbourhood
\thanks{Based on spectra collected at McDonald - USA, SAORAS - Russia,
 KPNO - USA, CTIO - Chile, MSO - Australia, OHP - France}}
\titlerunning{Galactic abundance gradient}
\author{S.M. Andrievsky,
\inst{1,2}\,
V.V. Kovtyukh,
\inst{2,3}\,
R.E. Luck,
\inst{4,5}\,
J.R.D. L\'epine,
 \inst{1}\, \\
D.  Bersier,
 \inst{6}\,
W.J. Maciel,
 \inst{1}\,
B. Barbuy,
 \inst{1}\,
V.G. Klochkova,
 \inst{7,8}\,
V.E. Panchuk,
 \inst{7,8}\,
R.U. Karpischek
 \inst{9}
}
\authorrunning{Andrievsky et al.}
\offprints{Andrievsky S.M.}
\institute{
Instituto Astron\^ {o}mico e Geof\' \i sico, Universidade de S\~ {a}o
Paulo, Av. Miguel Stefano, 4200, S\~ao Paulo SP, Brazil,
email: sergei@andromeda.iagusp.usp.br
\and
Department of Astronomy, Odessa State University,
Shevchenko Park, 65014, Odessa, Ukraine,
email: scan@deneb.odessa.ua; val@deneb.odessa.ua
\and
Odessa Astronomical Observatory and Isaac Newton Institute of Chile,
Odessa Branch, Ukraine
\and
Department of Astronomy, Case Western Reserve
University, 10900 Euclid Avenue, Cleveland, OH 44106-7215,
email: luck@fafnir.astr.cwru.edu
\and Visiting Astronomer, Cerro Tololo
Inter-American Observatory, National Optical Astronomy
Observatories which are operated by the Association of
Universities for Research in Astronomy, Inc., under contract with
the U.S. National Science Foundation
\and
Harvard-Smithsonian Center for Astrophysics, 60 Garden Street,
MS 16, Cambridge, MA 02138, USA,
email: dbersier@cfa.harvard.edu
\and
Special Astrophysical Observatory, Russian Academy of Sciences,
Nizhny Arkhyz, Stavropol Territory, 369167,Russia,
email: valenta@sao.ru; panchuk@sao.ru
\and
SAO RAS and Isaac Newton Institute of Chile, SAO RAS Branch, Russia
\and
EdIC group, Universidade de S\~ {a}o Paulo, S\~ao Paulo, Brazil,
email: ueda@ime.usp.be}
\date{Received ; accepted }
\abstract
{A number of studies of abundance gradients in the galactic disk have
been performed in recent years. The results obtained are rather disparate:
from no detectable gradient to a rather significant slope of about
$-0.1$ dex~kpc$^{-1}$. The present study concerns the abundance gradient based
on the spectroscopic analysis of a sample of classical Cepheids. These stars
enable one to obtain reliable abundances of a variety of chemical elements.
Additionally, they have well determined distances which allow an accurate
determination of abundance distributions in the galactic disc.
Using 236 high resolution spectra of 77 galactic Cepheids, the radial elemental
distribution in the galactic disc between galactocentric distances in the range
6-11 kpc has been investigated. Gradients for 25 chemical elements (from
carbon to gadolinium) are derived. The following results were obtained in this
study:
Almost all investigated elements show  rather flat abundance distributions in
the middle part of galactic disc. Typical values for iron-group elements
lie within an interval from $\approx -0.02$ to $\approx -0.04$ dex~kpc$^{-1}$
(in particular, for iron we obtained d[Fe/H]/dR$_{\rm G}= -0.029$ dex~kpc$^{-1}$).
Similar gradients were also obtained for O, Mg, Al, Si, and Ca.
For sulphur we have found a steeper gradient ($-0.05$ dex~kpc$^{-1}$).
For elements from Zr to Gd we obtained (within the error bars) a near to zero
gradient value. This result is reported for the first time.
Those elements whose abundance is not expected to be altered during the early
stellar evolution (e.g. the iron-group elements) show at the solar galactocentric
distance [El/H] values which are essentially solar. Therefore, there is no
apparent reason to consider our Sun as a metal-rich star.
The gradient values obtained in the present study indicate that the radial
abundance distribution within $\approx$ 10 kpc is quite homogeneous, and this
result favors a galactic model including a bar structure which may induce
radial flows in the disc, and thus may be responsible for abundance
homogenization.
}

\maketitle
\section{Introduction}

In recent years the problem of radial abundance gradients in spiral galaxies
has emerged as a central problem in the field of galactic chemodynamics.
Abundance gradients as observational characteristics of the galactic disc are
among the most important input parameters in any theory of galactic chemical
evolution. Further development of theories of galactic chemodynamics is
dramatically hampered by the scarcity of observational data, their large
uncertainties and, in some cases, apparent contradictions between independent
observational results. Many questions concerning the present-day abundance
distribution in the galactic disc, its spatial properties, and evolution with
time, still have to be answered.

Discussions of the galactic abundance gradient, as determined from several
studies, were provided by Friel (\cite{fr95}), Gummersbach et al. (\cite{guet98}),
Hou, Prantzos \& Boissier (\cite{hpb00}). Here we only briefly summarize some of the
more pertinent results.

1) A variety of objects (planetary nebulae, cool giants/supergiants, F-G dwarfs,
old open clusters) seem to give evidence that an abundance gradient
exists. Using DDO, Washington, UBV photometry and moderate resolution spectroscopy
combined with metallicity calibrations for open clusters and cool giants the
following gradients were derived (d[Fe/H]/dR$_{\rm G}$):
$-0.05$ dex~kpc$^{-1}$ (Janes \cite{jan79}),
$-0.095$ dex~kpc$^{-1}$ (Panagia \& Tosi \cite{pt81}),
$-0.07$ dex~kpc$^{-1}$ (Harris \cite{har81}),
$-0.11$ dex~kpc$^{-1}$ (Cameron \cite{cam85}),
$-0.017$ dex~kpc$^{-1}$ (Neese \& Yoss \cite{ny88}),
$-0.13$ dex~kpc$^{-1}$ (Geisler, Clari\'a \& Minniti \cite{gcm92}),
$-0.097$ dex~kpc$^{-1}$ (Thogersen, Friel \& Fallon \cite{tff93}),
$-0.09$ dex~kpc$^{-1}$ (Friel \& Janes \cite{fj93}),
$-0.091$ dex~kpc$^{-1}$ (Friel \cite{fr95}),
$-0.09$ dex~kpc$^{-1}$ (Carraro, Ng \& Portinary \cite{cnp98}),
$-0.06$ dex~kpc$^{-1}$ (Friel \cite{fr99}, Phelps \cite{ph00}).

One must also add that there have been attempts to derive the abundance gradient
(specifically d[Fe/H]/dR$_{\rm G}$) using high-resolution spectroscopy of cool
giant and supergiant stars. Harris \& Pilachowski (\cite{hp84}) obtained
$-0.07$ dex~kpc$^{-1}$, while Luck (\cite{luck82}) found a steeper gradient of
$-0.13$ dex~kpc$^{-1}$.

Oxygen and sulphur gradients determined from observations of planetary nebulae
are $-0.058$ dex~kpc$^{-1}$ and  $-0.077$ dex~kpc$^{-1}$ respectively
(Maciel \& Quireza \cite{mq99}), with slightly flatter values for neon and argon,
as in Maciel \& K\"oppen (\cite{mk94}). A smaller slope was found in an earlier study of
Pasquali \& Perinotto (\cite{pp93}). According to those authors the nitrogen abundance
gradient is $-0.052$ dex~kpc$^{-1}$, while that of oxygen is $-0.030$ dex~kpc$^{-1}$.

2) From young B main sequence stars, Smartt \& Rolleston (\cite{sr97}) found
a gradient of $-0.07$ dex~kpc$^{-1}$, while Gehren et al. (\cite{gehet85}),
Fitzsimmons, Dufton \& Rolleston (\cite{fdr92}), Kaufer et al. (\cite{kaet94})
and Kilian-Montenbruck, Gehren \& Nissen (\cite{kmgn94}) derived significantly
smaller values: $-0.03-0.00$ dex~kpc$^{-1}$. No systematic abundance
variation with galactocentric distance was found by Fitzsimmons et al.
(\cite{fitet90}). The recent studies of Gummersbach et al. (\cite{guet98})
and Rolleston et al. (\cite{rollet00}) support the existence of a gradient
($-0.07$ dex~kpc$^{-1}$). The elements in these studies were C-N-O and Mg-Al-Si.

3) Studies of the abundance gradient (primarily nitrogen, oxygen, sulphur) in
the Galactic disc based on young objects such as \ion{H}{ii} regions give
positive results: either significant slopes from $-0.07$ to $-0.11$
dex~kpc$^{-1}$ according to: Shaver et al. (\cite{shavet83}) for nitrogen
and oxygen, Simpson et al. (\cite{simpet95}) for nitrogen and sulphur, Afflerbach,
Churchwell \& Werner (\cite{afflet97}) for nitrogen, Rudolph et al. (\cite{rudet97})
for nitrogen and sulphur, or intermediate gradients of about $-0.05$ to $-0.06$ dex~kpc$^{-1}$
according to: Simpson \& Rubin (\cite{simru90}) for sulphur, Afflerbach, Churchwell
\& Werner (\cite{afflet97}) for oxygen and sulphur; and negative ones: weak or
nonexistent gradients as concluded by Fich \& Silkey \cite{fs91}; Vilchez \&
Esteban \cite{viles96}, Rodriguez (\cite{rod99}).
Recently Pe\~na et al. (\cite{pet00}) derived oxygen abundances in several
\ion{H}{ii} regions and found a rather flat distribution with galactocentric
distance (coefficient $-0.04$ dex~kpc$^{-1}$). The same results were also reported
by Deharveng et al. (\cite{deet00}).

As one can see, there is no conclusive argument allowing one to come to a
definite conclusion about whether or not a significant abundance gradient
exists in the galactic disc, at least for all elements considered and
within the whole observed interval of galactocentric distances.
Compared to other objects supplying us with an information about the radial
distribution of elemental abundances in the galactic disc, Cepheids have
several advantages:

1) they are primary distance calibrators which provide excellent distance
estimates;

2) they are luminous stars allowing one to probe to large distances;

3) the abundances of many chemical elements can be measured from Cepheid
spectra (many more than from \ion{H}{ii} regions or B stars). This is important
for investigation of the distribution in the galactic disc of absolute abundances
and abundance ratios.  Additionally, Cepheids allow the study of abundances past
the iron-peak which  are not generally available in \ion{H}{ii} regions
or B stars;

4) lines in Cepheid spectra are sharp and well-defined which enables one to
derive elemental abundances with high reliability.

In view of the inconsistencies in the current results on the galactic
abundance gradient, and those advantages which are afforded by Cepheids, we
have undertaken a large survey of Cepheids in order to provide independent
information which should be useful as boundary conditions for theories
of galactic chemodynamics. We also hope that the results on the abundance
gradient from the Cepheids will also be helpful to constrain the structure
and age of the bar, and its influence on the metallicity gradient.
This first paper in this series on abundance gradients from Cepheids presents
the results for the solar neighbourhood.

\section{Observations}

For the great majority of the program stars multiphase observations were
obtained. From the total number of the spectra for each star we selected
those showing no or at most a small asymmetry of the spectral lines. For the
distant (fainter) Cepheids we have analyzed 3-4 spectra in order to derive
the abundances, while for the nearby stars 2-3 spectra were used.
This is predicated on the fact that the brighter stars have higher S/N spectra
and thus better determined equivalent widths. For some stars we have only one
spectrum, and for a few Cepheids more than four spectra were analyzed.

Information about the program stars and spectra is given in Table 1. Note that
we also added to our sample two distant Cepheids (TV Cam and YZ Aur) which were
previously analyzed by Harris \& Pilachowski (\cite{hp84}). We have used their data
for these stars but atmospheric parameters and elemental abundances (specifically
the iron content) were re-determined using the same methodology as for other
program stars (see next Section).

\begin{table*}
\caption[]{Program Cepheids, their spectra and results for individual phases}
\tiny
\begin{tabular}{ccccccccc}
\hline
Star          & P, d   & JD, 24+     &$\phi$ & Telescope &T$_{\rm eff},\,K$&
$\log g$&V$_{\rm t}$,\,km~s$^{-1}$&[Fe/H]\\
\hline
\object{V473 Lyr} (s)  & 1.4908 &  49906.43160&  0.793& OHP 1.93m& 6163 & 2.45 & 4.20&--0.09 \\
              &        &  49907.57360&  0.559& OHP 1.93m& 6113 & 2.60 & 4.50&--0.05 \\
              &        &             &       &          &      &      &     &       \\
\object{SU Cas}(s)     & 1.9493 &  50674.95633&  0.902& MDO 2.1m & 6594 & 2.60 & 3.85&--0.02 \\
              &        &  50675.96550&  0.420& MDO 2.1m & 6162 & 2.25 & 3.00&--0.00 \\
              &        &  50678.93059&  0.941& MDO 2.1m & 6603 & 2.50 & 3.50&--0.01 \\
              &        &  51473.79052&  0.704& MDO 2.1m & 6201 & 2.30 & 2.85&--0.00 \\
              &        &             &       &          &      &      &     &       \\
\object{EU Tau}   (s)  & 2.1025 &  51096.90943&  0.172& MDO 2.1m & 6203 & 2.00 & 3.00&--0.09 \\
              &        &  51097.89587&  0.641& MDO 2.1m & 6014 & 2.20 & 3.30&--0.03 \\
              &        &             &       &          &      &      &     &       \\
\object{IR Cep}   (s)  & 2.1140 &  48821.46940&  0.137& SAORAS 6m& 6162 & 2.40 & 4.10& +0.00 \\
              &        &             &       &          &      &      &     &       \\
\object{TU Cas}        & 2.1393 &  50674.90819&  0.919& MDO 2.1m & 6465 & 2.10 & 3.70&--0.07 \\
              &        &  50675.91792&  0.391& MDO 2.1m & 5993 & 2.40 & 4.00& +0.03 \\
              &        &  50677.89839&  0.317& MDO 2.1m & 6013 & 2.30 & 3.00& +0.07 \\
              &        &  50678.91008&  0.790& MDO 2.1m & 6148 & 2.40 & 4.80&--0.05 \\
              &        &  50735.76094&  0.364& MDO 2.1m & 5905 & 2.10 & 2.60& +0.09 \\
              &        &  50739.76484&  0.235& MDO 2.1m & 6185 & 2.40 & 3.50& +0.05 \\
              &        &  50740.77928&  0.710& MDO 2.1m & 5906 & 2.20 & 3.90& +0.08 \\
              &        &  50741.79234&  0.183& MDO 2.1m & 6026 & 1.80 & 4.10&--0.06 \\
              &        &  51053.90600&  0.076& MDO 2.1m & 6520 & 2.30 & 3.50&--0.03 \\
              &        &  51095.78573&  0.649& MDO 2.1m & 5920 & 2.30 & 4.00& +0.09 \\
              &        &  51097.81206&  0.600& MDO 2.1m & 5857 & 2.20 & 4.50& +0.08 \\
              &        &             &       &          &      &      &     &       \\
\object{DT Cyg}   (s)  & 2.4991 &  50379.65482&  0.014& MDO 2.1m & 6406 & 2.60 & 3.70& +0.14 \\
              &        &  50383.55796&  0.576& MDO 2.1m & 6010 & 2.30 & 3.50& +0.08 \\
              &        &  50674.86130&  0.140& MDO 2.1m & 6384 & 2.40 & 3.50& +0.08 \\
              &        &             &       &          &      &      &     &       \\
\object{V526 Mon}  (s) & 2.6750 &  49022.42220&  0.674& SAORAS 6m& 6464 & 2.40 & 3.50&--0.13 \\
              &        &             &       &          &      &      &     &       \\
\object{V351 Cep}  (s) & 2.8060 &  48853.50600&  0.425& SAORAS 6m& 5944 & 2.50 & 4.30& +0.03 \\
              &        &  49203.37300&  0.301& SAORAS 6m& 6005 & 2.10 & 3.30& +0.02 \\
              &        &             &       &          &      &      &     &       \\
\object{VX Pup}        & 3.0109 &  51231.52248&  0.813& MSO 74in & 6159 & 2.50 & 3.50& -0.12 \\
              &        &             &       &          &      &      &     &       \\
\object{SZ Tau}(s)     & 3.1484 &  50379.86757&  0.430& MDO 2.1m & 5901 & 2.10 & 3.30& +0.12 \\
              &        &  50380.89275&  0.756& MDO 2.1m & 5955 & 2.30 & 3.90& +0.06 \\
              &        &  50482.66323&  0.077& MDO 2.1m & 6121 & 2.20 & 3.70& +0.03 \\
              &        &             &       &          &      &      &     &       \\
\object{V1334 Cyg} (s) & 3.3330 &  50676.84617&  0.185& MDO 2.1m & 6149 & 1.90 & 3.50&--0.02 \\
              &        &  50738.72218&  0.751& MDO 2.1m & 6363 & 2.00 & 3.60&--0.05 \\
              &        &  51093.72224&  0.268& MDO 2.1m & 6210 & 2.20 & 3.90&--0.02 \\
              &        &             &       &          &      &      &     &       \\
\object{BG Cru} (s)    & 3.3427 &  51231.64722&  0.188& MSO 74in & 6101 & 2.00 & 3.80&--0.02 \\
              &        &             &       &          &      &      &     &       \\
\object{BD Cas} (s)    & 3.6510 &  49572.49400&  0.772& SAORAS 6m& 6200 & 2.50 & 5.00&--0.11 \\
              &        &  49577.33900&  0.073& SAORAS 6m& 6075 & 2.30 & 4.50&--0.12 \\
              &        &  49578.34300&  0.347& SAORAS 6m& 5880 & 2.40 & 4.00&--0.04 \\
              &        &             &       &          &      &      &     &       \\
\object{RT Aur}        & 3.7282 &  50736.01017&  0.328& MDO 2.1m & 5982 & 1.90 & 3.00& +0.03 \\
              &        &  50736.96014&  0.583& MDO 2.1m & 5686 & 1.85 & 3.40& +0.07 \\
              &        &  50739.91812&  0.377& MDO 2.1m & 5878 & 2.00 & 3.00& +0.08 \\
              &        &             &       &          &      &      &     &       \\
\object{DF Cas}        & 3.8320 &  50505.18472&  0.401& SAORAS 6m& 5644 & 2.20 & 4.65& +0.13 \\
              &        &             &       &          &      &      &     &       \\
\object{SU Cyg}        & 3.8455 &  50736.68203&  0.415& MDO 2.1m & 5956 & 2.10 & 3.20&--0.00 \\
              &        &  50738.70189&  0.940& MDO 2.1m & 6314 & 2.40 & 4.50&--0.03 \\
              &        &             &       &          &      &      &     &       \\
\object{ST Tau}        & 4.0325 &  51096.86237&  0.869& MDO 2.1m & 6519 & 2.50 & 4.40&--0.02 \\
              &        &  51097.92470&  0.132& MDO 2.1m & 6268 & 2.00 & 3.50&--0.05 \\
              &        &  51474.97888&  0.594& MDO 2.1m & 5676 & 1.80 & 3.90&--0.09 \\
              &        &             &       &          &      &      &     &       \\
\object{V1726 Cyg}  (s) & 4.2360 &  51003.23100&  0.109& SAORAS 6m& 6349 & 2.20 & 5.20&--0.02 \\
              &        &             &       &          &      &      &     &       \\
\object{BQ Ser}        & 4.2709 &  51659.96271&  0.121& MDO 2.1m & 6296 & 2.20 & 4.00&--0.10 \\
              &        &  51660.96355&  0.355& MDO 2.1m & 6021 & 1.70 & 3.15&--0.02 \\
              &        &  51661.96295&  0.589& MDO 2.1m & 5768 & 1.80 & 3.20&--0.00 \\
              &        &             &       &          &      &      &     &       \\
\hline
\end{tabular}
\begin{itemize}
\item[] MDO 2.1m - McDonald Observatory (USA), Struve 2.1-m reflector, R = 60,000,
S/N $>$ 100.
\item[] KPNO 4m - Kitt Peak National Observatory (USA), 4-m and coud\'{e}-feed
telescope, R = 30,000 and 80,000 respectively, S/N $\approx$150 (except for
CV Mon with a S/N of about 50).
\item[] CTIO 4m - Cerro Tololo Inter-American Observatory (Chile), 4-m telscope,
R = 30,000, S/N $>$ 100.
\item[] MSO 74in - Mount Stromlo Observatory (Australia), 74-inch telescope,
R = 56,000, S/N $\le$ 50.
\item[] SAORAS 6m - Special Astrophysical Observatory of the Russian Academy of
Sciences (Russia), 6-m telescope, R = 14,000 and 25,000, S/N $\approx$ 70-100
(except for DF Cas, V924 Cyg and TX Del, where the S/N is below 70).
\item[] OHP 1.93m - Haute-Provence Observatoire (France), 1.93-m telescope,
R = 40,000 , S/N $>$ 150.
\end{itemize}
\end{table*}

\begin{table*}
{\bf Table 1 (continued)}\\
\tiny
\begin{tabular}{ccccccccc}
\hline
Star          & P, d   & JD, 24+     &$\phi$ & Telescope &T$_{\rm eff}$,\,K&$\log g$
&V$_{\rm t}$,\,km~s$^{-1}$&[Fe/H]\\
\hline
\object{Y  Lac}        & 4.2338 &  51098.78394&  0.936& MDO 2.1m & 6330 & 2.00 & 4.00&--0.10 \\
              &        &  51474.61215&  0.858& MDO 2.1m & 6006 & 1.70 & 4.45&--0.08 \\
              &        &  51475.67351&  0.103& MDO 2.1m & 6258 & 1.80 & 4.00&--0.09 \\
              &        &             &       &          &      &      &     &       \\
\object{T Vul}& 4.4355 &  50381.66164&  0.144& MDO 2.1m & 6077 & 2.00 & 3.55& +0.02 \\
              &        &  50382.67995&  0.374& MDO 2.1m & 5768 & 2.00 & 3.60& +0.03 \\
              &        &  51095.55920&  0.094& MDO 2.1m & 6174 & 2.00 & 3.70&--0.01 \\
              &        &             &       &          &      &      &     &       \\
\object{FF Aql}  (s)  & 4.4709 &  50674.68472&  0.987& MDO 2.1m & 6425 & 2.10 & 4.90&--0.02 \\
              &        &  50677.74557&  0.672& MDO 2.1m & 6083 & 2.00 & 4.80& +0.05 \\
              &        &  50678.71213&  0.888& MDO 2.1m & 6421 & 2.10 & 5.40& +0.00 \\
\object{CF Cas}    & 4.8752 &  50735.81552&  0.980& MDO 2.1m & 6115 & 2.00 & 4.00&--0.03 \\
          &        &  50738.76114&  0.584& MDO 2.1m & 5454 & 1.70 & 4.30& +0.01 \\
          &        &  51055.92310&  0.641& MDO 2.1m & 5439 & 1.70 & 4.40&--0.01 \\
          &        &  51097.83265&  0.238& MDO 2.1m & 5704 & 1.90 & 3.70& +0.02 \\
          &        &  51098.85759&  0.448& MDO 2.1m & 5428 & 1.30 & 3.40&--0.01 \\
          &        &             &       &          &      &      &     &       \\
\object{TV Cam}    & 5.2950 &  44333.65000&  0.090& KPNO 4m  & 6049 & 2.15 & 4.30&--0.06 \\
          &        &             &       &          &      &      &     &       \\
\object{BG Lac}    & 5.3319 &  51055.83161&  0.147& MDO 2.1m & 5923 & 1.90 & 3.80& +0.01 \\
          &        &  51056.81572&  0.332& MDO 2.1m & 5625 & 1.85 & 3.60& +0.02 \\
          &        &  51097.77450&  0.014& MDO 2.1m & 6095 & 1.80 & 4.20&--0.06 \\
          &        &             &       &          &      &      &     &       \\
\object{$\delta$ Cep}& 5.3663&  50379.68413&  0.561&MDO 2.1m & 5544 & 1.70 & 3.70& +0.08 \\
          &        &  50741.70985&  0.024& MDO 2.1m & 6532 & 2.20 & 4.10& +0.05 \\
          &        &             &       &          &      &      &     &       \\
\object{V1162 Aql} & 5.3760 &  51774.74747&  0.890& MDO 2.1m & 5989 & 2.00 & 4.60&--0.03 \\
          &        &  51775.64945&  0.058& MDO 2.1m & 5940 & 1.80 & 3.90& +0.04 \\
          &        &             &       &          &      &      &     &       \\
\object{CV Mon}    & 5.3789 &  48878.97917&  0.178& KPNO 4m  & 5897 & 2.00 & 4.10&--0.03 \\
          &        &             &       &          &      &      &     &       \\
\object{V Cen}     & 5.4939 &  49116.54792&  0.236& CTIO 4m  & 5705 & 2.10 & 3.90& +0.04 \\
          &        &             &       &          &      &      &     &       \\
\object{V924 Cyg} (s:)& 5.5710&  48819.4062& 0.702& SAORAS 6m& 5910 & 1.80 & 5.00&--0.09 \\
          &        &             &       &          &      &      &     &       \\
\object{MY Pup} (s)& 5.6953 &  51231.46954&  0.950& MSO 74in & 6170 & 1.85 & 3.30&--0.12 \\
          &        &             &       &          &      &      &     &       \\
\object{Y Sgr}    & 5.7734 &  50674.62981&  0.911& MDO 2.1m & 6078 & 1.90 & 4.40& +0.07 \\
          &        &  51053.66761&  0.564& MDO 2.1m & 5490 & 1.60 & 3.90& +0.06 \\
          &        &  51057.67128&  0.257& MDO 2.1m & 5841 & 1.55 & 3.90& +0.05 \\
          &        &             &       &          &      &      &     &       \\
\object{EW Sct}    & 5.8233 &  51053.69933&  0.722& MDO 2.1m & 5728 & 1.80 & 3.50& +0.05 \\
          &        &  51055.68163&  0.062& MDO 2.1m & 6155 & 2.30 & 5.00& +0.01 \\
          &        &  51058.65288&  0.572& MDO 2.1m & 5655 & 1.70 & 3.40& +0.05 \\
          &        &             &       &          &      &      &     &       \\
\object{FM Aql}    & 6.1142 &  50736.61650&  0.942& MDO 2.1m & 6255 & 1.80 & 4.10& +0.07 \\
          &        &  50738.60784&  0.267& MDO 2.1m & 5750 & 1.50 & 3.50& +0.11 \\
          &        &             &       &          &      &      &     &       \\
\object{TX Del}    & 6.1660 &  49165.12300&  0.005& SAORAS 6m& 6217 & 1.80 & 6.00& +0.23 \\
          &        &             &       &          &      &      &     &       \\
\object{V367 Sct}  & 6.2931 &  51003.39510&       & SAORAS 6m& 5891 & 2.10 & 4.25& --0.01\\
          &        &             &       &          &      &      &     &       \\
\object{X Vul}    & 6.3195 &  50738.66499&  0.407& MDO 2.1m & 5649 & 1.80 & 3.45& +0.09 \\
          &        &  50739.67020&  0.566& MDO 2.1m & 5434 & 1.60 & 4.00& +0.05 \\
          &        &  51097.69899&  0.220& MDO 2.1m & 5875 & 1.80 & 4.10& +0.09 \\
          &        &             &       &          &      &      &     &       \\
\object{AW Per}    & 6.4636 &  50380.92178&  0.994& MDO 2.1m & 6423 & 2.15 & 4.30&--0.03 \\
          &        &  50382.86865&  0.295& MDO 2.1m & 5989 & 1.90 & 3.60& +0.06 \\
          &        &  50383.87798&  0.451& MDO 2.1m & 5836 & 2.00 & 3.70& +0.11 \\
          &        &  50736.87128&  0.065& MDO 2.1m & 6627 & 1.70 & 3.90&--0.06 \\
          &        &             &       &          &      &      &     &       \\
\object{U Sgr}     & 6.7452 &  50674.63784&  0.550& MDO 2.1m & 5388 & 1.50 & 4.00& +0.06 \\
          &        &  50674.64319&  0.551& MDO 2.1m & 5416 & 1.70 & 4.00& +0.07 \\
          &        &  50677.67816&  0.001& MDO 2.1m & 6145 & 1.90 & 4.70& +0.01 \\
          &        &  50735.55199&  0.581& MDO 2.1m & 5347 & 1.60 & 4.00& +0.04 \\
          &        &  50736.56445&  0.731& MDO 2.1m & 5399 & 1.70 & 5.20& +0.01 \\
          &        &  50739.57384&  0.178& MDO 2.1m & 5876 & 1.70 & 4.00& +0.08 \\
          &        &  50740.57748&  0.326& MDO 2.1m & 5710 & 1.70 & 4.00& +0.09 \\
          &        &  50741.55532&  0.471& MDO 2.1m & 5475 & 1.60 & 4.00& +0.05 \\
          &        &  50949.66389&  0.326& MDO 2.1m & 5705 & 1.70 & 4.00& +0.05 \\
          &        &  51053.67960&  0.746& MDO 2.1m & 5441 & 1.80 & 5.50& +0.01 \\
          &        &  51054.68627&  0.896& MDO 2.1m & 6077 & 2.10 & 5.50& +0.04 \\
          &        &  51094.62511&  0.817& MDO 2.1m & 5746 & 2.00 & 6.00& +0.02 \\
          &        &             &       &          &      &      &     &       \\
\object{V496 Aql} (s)& 6.8071&  51774.72348& 0.910& MDO 2.1m & 5822 & 1.70 & 4.25& +0.03 \\
          &        &  51775.62763&  0.043& MDO 2.1m & 5841 & 1.70 & 4.00& +0.06 \\
          &        &             &       &          &      &      &     &       \\
\object{$\eta$ Aql}& 7.1767 &  50739.66527&  0.021& MDO 2.1m & 6275 & 1.90 & 4.40& +0.04 \\
          &        &  50741.62832&  0.295& MDO 2.1m & 5787 & 1.80 & 3.90& +0.06 \\
          &        &             &       &          &      &      &     &       \\
\hline
\end{tabular}
\end{table*}

\begin{table*}
{\bf Table 1 (continued)}\\
\tiny
\begin{tabular}{ccccccccc}
\hline
Star          & P, d   & JD, 24+     &$\phi$ & Telescope &T$_{\rm eff}$,\,K
&$\log g$&V$_{\rm t}$,\,km~s$^{-1}$&[Fe/H]\\
\hline
\object{BB Her}    & 7.5080 &  51055.64149&  0.665& MDO 2.1m & 5265 & 1.60 & 4.10& +0.08 \\
          &        &  51058.66436&  0.068& MDO 2.1m & 5988 & 1.80 & 4.30& +0.17 \\
          &        &  51097.63866&  0.259& MDO 2.1m & 5750 & 1.80 & 4.20& +0.16 \\
          &        &  51098.61695&  0.389& MDO 2.1m & 5556 & 1.70 & 4.20& +0.13 \\
          &        &             &       &          &      &      &     &       \\
\object{RS Ori}    & 7.5669 &  51098.98467&  0.012& MDO 2.1m & 6367 & 1.80 & 3.70&--0.10 \\
          &        &  51476.87338&  0.952& MDO 2.1m & 6666 & 2.10 & 4.40&--0.10 \\
          &        &  51568.74952&  0.094& MDO 2.1m & 6193 & 1.70 & 3.90&--0.12 \\
          &        &  51569.75694&  0.228& MDO 2.1m & 6043 & 1.60 & 3.70&--0.07 \\
\object{V440 Per} (s)& 7.5700 &  50738.81027&  0.280& MDO 2.1m & 6144 & 2.00 & 5.10&--0.03 \\
          &        &  50741.86515&  0.684& MDO 2.1m & 5997 & 1.90 & 4.90&--0.08 \\
          &        &  51098.90018&  0.848& MDO 2.1m & 6021 & 1.85 & 5.20&--0.06 \\
          &        &             &       &          &      &      &     &       \\
\object{W Sgr}    & 7.5949 &  50741.55035&  0.045& MDO 2.1m & 6446 & 2.10 & 4.60&--0.04 \\
          &        &  51053.62870&  0.137& MDO 2.1m & 6207 & 2.00 & 4.30& +0.02 \\
          &        &  51056.63553&  0.533& MDO 2.1m & 5540 & 1.65 & 3.80& +0.01 \\
          &        &             &       &          &      &      &     &       \\
\object{RX Cam}    & 7.9120 &  50735.89188&  0.212& MDO 2.1m & 5942 & 1.95 & 4.10& +0.07 \\
          &        &  50736.84355&  0.332& MDO 2.1m & 5755 & 1.80 & 4.10& +0.06 \\
          &        &  50741.88410&  0.969& MDO 2.1m & 6227 & 1.90 & 4.20&--0.02 \\
          &        &             &       &          &      &      &     &       \\
\object{W Gem}    & 7.9138 &  51096.95531&  0.009& MDO 2.1m & 6003 & 1.85 & 4.05&--0.02 \\
          &        &  51097.96579&  0.137& MDO 2.1m & 6021 & 1.80 & 3.90&--0.08 \\
          &        &  51098.96898&  0.264& MDO 2.1m & 5704 & 1.90 & 4.50&--0.01 \\
          &        &             &       &          &      &      &     &       \\
\object{U Vul}    & 7.9906 &  51055.76933&  0.415& MDO 2.1m & 5629 & 1.70 & 4.00& +0.09 \\
          &        &  51056.73122&  0.536& MDO 2.1m & 5490 & 1.50 & 3.70& +0.08 \\
          &        &  51475.65134&  0.961& MDO 2.1m & 6314 & 1.90 & 5.00& +0.01 \\
          &        &             &       &          &      &      &     &       \\
\object{DL Cas}    & 8.0007 &  50381.76574&  0.105& MDO 2.1m & 5860 & 1.70 & 4.70&--0.02 \\
          &        &  50382.75636&  0.229& MDO 2.1m & 5786 & 1.70 & 4.20& +0.02 \\
          &        &  50736.74222&  0.473& MDO 2.1m & 5438 & 1.40 & 4.00&--0.05 \\
          &        &             &       &          &      &      &     &       \\
\object{AC Mon}    & 8.0143 &  50505.35490&  0.919& SAORAS 6m& 6121 & 2.20 & 5.80&--0.07 \\
          &        &             &       &          &      &      &     &       \\
\object{V636 Cas} (s)& 8.3770& 50735.78055&  0.919& MDO 2.1m & 5562 & 1.50 & 4.10& +0.05 \\
          &        &  50736.77703&  0.038& MDO 2.1m & 5473 & 1.50 & 3.80& +0.06 \\
          &        &  50737.79151&  0.159& MDO 2.1m & 5395 & 1.60 & 4.05& +0.07 \\
          &        &             &       &          &      &      &     &       \\
\object{S Sge}    & 8.3821 &  50675.76850&  0.051& MDO 2.1m & 6135 & 2.00 & 4.30& +0.12 \\
          &        &  50677.81371&  0.295& MDO 2.1m & 5855 & 1.80 & 3.80& +0.11 \\
          &        &  50741.65318&  0.911& MDO 2.1m & 6093 & 2.10 & 4.95& +0.09 \\
          &        &             &       &          &      &      &     &       \\
\object{GQ Ori}    & 8.6161 &  49022.30620&  0.365& SAORAS 6m& 5732 & 1.75 & 5.20&--0.03 \\
          &        &             &       &          &      &      &     &       \\
\object{V500 Sco}  & 9.3168 &  51093.57083&  0.340& MDO 2.1m & 5359 & 1.40 & 3.80&--0.03 \\
          &        &  51094.56998&  0.447& MDO 2.1m & 5243 & 1.40 & 3.80&--0.05 \\
          &        &  51097.55554&  0.768& MDO 2.1m & 6050 & 1.80 & 4.00& +0.01 \\
          &        &  51098.55823&  0.875& MDO 2.1m & 5969 & 1.70 & 4.40&--0.03 \\
          &        &             &       &          &      &      &     &       \\
\object{FN Aql}    & 9.4816 &  50740.65730&  0.794& MDO 2.1m & 5698 & 1.65 & 4.70&--0.05 \\
          &        &  51055.73047&  0.024& MDO 2.1m & 5922 & 1.90 & 4.90&--0.02 \\
          &        &  51056.70025&  0.126& MDO 2.1m & 5729 & 1.55 & 3.70& +0.00 \\
          &        &  51057.74039&  0.236& MDO 2.1m & 5464 & 1.50 & 3.50& +0.01 \\
          &        &             &       &          &      &      &     &       \\
\object{YZ Sgr}    & 9.5536 &  50735.57390&  0.249& MDO 2.1m & 5496 & 1.30 & 3.65& +0.04 \\
          &        &  50737.54744&  0.456& MDO 2.1m & 5223 & 1.20 & 3.90& +0.08 \\
          &        &  50740.59629&  0.775& MDO 2.1m & 5906 & 1.80 & 4.80& +0.06 \\
          &        &  51057.70299&  0.967& MDO 2.1m & 5943 & 1.50 & 4.10& +0.05 \\
          &        &             &       &          &      &      &     &       \\
\object{S Nor}     & 9.7542 &  49116.83715&  0.550& CTIO 4m  & 5797 & 2.00 & 4.60& +0.06 \\
          &        &  49497.47640&  0.570& CTIO 4m  & 5677 & 1.90 & 5.80& +0.06 \\
          &        &             &       &          &      &      &     &       \\
\object{$\beta$ Dor}& 9.8424 &  51231.44353& 0.128& MSO 74in & 5618 & 1.60 & 4.30&--0.01 \\
          &        &             &       &          &      &      &     &       \\
\object{$\zeta$ Gem}&10.1507 &  50381.98995&  0.815& MDO 2.1m & 5740 & 1.70 & 4.50& +0.01 \\
          &        &  50736.98149&  0.787& MDO 2.1m & 5741 & 1.70 & 4.50& +0.03 \\
          &        &  50739.96631&  0.081& MDO 2.1m & 5593 & 1.40 & 3.50& +0.06 \\
          &        &             &       &          &      &      &     &       \\
\object{Z Lac}    &10.8856 &  51055.88242&  0.903& MDO 2.1m & 5899 & 1.70 & 4.30& +0.01 \\
          &        &  51056.86659&  0.993& MDO 2.1m & 6432 & 1.90 & 4.50&--0.03 \\
          &        &  51058.86037&  0.177& MDO 2.1m & 5722 & 1.50 & 3.80& +0.05 \\
          &        &  51093.78777&  0.385& MDO 2.1m & 5241 & 1.20 & 3.30& +0.02 \\
          &        &             &       &          &      &      &     &       \\
\object{VX Per}    &10.8890 &  50739.78515&  0.084& MDO 2.1m & 5515 & 1.40 & 3.80&--0.03 \\
          &        &  50740.81077&  0.178& MDO 2.1m & 5369 & 1.40 & 3.80&--0.03 \\
          &        &  51095.81172&  0.780& MDO 2.1m & 5989 & 1.70 & 4.20&--0.07 \\
          &        &  51096.77199&  0.868& MDO 2.1m & 6026 & 1.70 & 4.20&--0.05 \\
          &        &             &       &          &      &      &     &       \\
\object{V340 Nor}(s:)&11.2870& 49116.86597&  0.193& CTIO 4m  & 5595 & 1.75 & 4.50& +0.00 \\
          &        &             &       &          &      &      &     &       \\
\hline
\end{tabular}
\end{table*}

\begin{table*}
{\bf Table 1 (continued)}\\
\tiny
\begin{tabular}{ccccccccc}
\hline
Star          & P, d   & JD, 24+     &$\phi$ & Telescope &T$_{\rm eff}$,\,K
&$\log g$&V$_{\rm t}$,\,km~s$^{-1}$&[Fe/H]\\
\hline
\object{RX Aur}    &11.6235 &  50736.91146&  0.171& MDO 2.1m & 5856 & 1.65 & 4.10&--0.03 \\
          &        &  50737.90348&  0.257& MDO 2.1m & 5677 & 1.45 & 3.70&--0.06 \\
          &        &  51093.93187&  0.885& MDO 2.1m & 6111 & 1.60 & 4.40&--0.08 \\
          &        &  51094.91699&  0.970& MDO 2.1m & 6312 & 1.70 & 4.30&--0.11 \\
\object{TT Aql}    &13.7547 &  51058.71730&  0.895& MDO 2.1m & 5630 & 1.65 & 5.10& +0.09 \\
          &        &  51093.61314&  0.432& MDO 2.1m & 5080 & 1.10 & 3.60& +0.12 \\
          &        &  51476.61856&  0.276& MDO 2.1m & 5335 & 1.15 & 3.60& +0.12 \\
          &        &             &       &          &      &      &     &       \\
\object{SV Mon}    &15.2328 &  50735.97014&  0.704& MDO 2.1m & 4916 & 1.05 & 4.70&--0.09 \\
          &        &  50737.92962&  0.832& MDO 2.1m & 5263 & 1.40 & 6.80&--0.06 \\
          &        &  50739.94097&  0.964& MDO 2.1m & 5482 & 1.40 & 4.90&--0.06 \\
          &        &  50741.93497&  0.095& MDO 2.1m & 6141 & 1.50 & 4.60&--0.03 \\
          &        &             &       &          &      &      &     &       \\
\object{X Cyg}    &16.3863 &  51056.80718&  0.020& MDO 2.1m & 6039 & 1.65 & 4.70& +0.12 \\
          &        &  51057.83565&  0.083& MDO 2.1m & 5741 & 1.5  & 4.20& +0.13 \\
          &        &             &       &          &      &      &     &       \\
\object{RW Cam}    &16.4148 &  50737.83335&  0.182& MDO 2.1m & 5368 & 1.15 & 3.60& +0.07 \\
          &        &  50738.83656&  0.243& MDO 2.1m & 5227 & 1.05 & 3.70& +0.01 \\
          &        &  50739.83773&  0.304& MDO 2.1m & 5108 & 0.95 & 3.40& +0.02 \\
          &        &  51095.85239&  0.993& MDO 2.1m & 5969 & 1.85 & 5.00& +0.03 \\
          &        &             &       &          &      &      &     &       \\
\object{CD Cyg}    &17.0740 &  50676.85264&  0.943& MDO 2.1m & 5484 & 1.45 & 5.20& +0.09 \\
          &        &  50677.84369&  0.001& MDO 2.1m & 6490 & 1.95 & 6.70& +0.01 \\
          &        &  50736.69273&  0.448& MDO 2.1m & 5108 & 1.10 & 4.00& +0.15 \\
          &        &             &       &          &      &      &     &       \\
\object{Y Oph} (s)&17.1269 &  50674.65908&  0.936& MDO 2.1m & 6013 & 1.75 & 5.60& +0.02 \\
          &        &  50736.55133&  0.551& MDO 2.1m & 5561 & 1.80 & 5.40& +0.05 \\
          &        &  50740.55568&  0.784& MDO 2.1m & 5709 & 1.90 & 5.80& +0.06 \\
          &        &  51054.66483&  0.128& MDO 2.1m & 6153 & 1.90 & 5.80& +0.07 \\
          &        &             &       &          &      &      &     &       \\
\object{SZ Aql}    &17.1408 &  50735.60696&  0.055& MDO 2.1m & 5568 & 1.20 & 4.30& +0.18 \\
          &        &  50737.65009&  0.174& MDO 2.1m & 5240 & 1.00 & 3.80& +0.13 \\
          &        &  51055.69289&  0.722& MDO 2.1m & 5454 & 1.45 & 6.00& +0.15 \\
          &        &  51057.71991&  0.840& MDO 2.1m & 6559 & 2.00 & 6.00& +0.13 \\
          &        &             &       &          &      &      &     &       \\
\object{YZ Aur}    &18.1932 &  44332.70000&  0.390& KPNO 4m  & 5175 & 1.65 & 4.70&--0.05 \\
          &        &             &       &          &      &      &     &       \\
\object{WZ Sgr}    &21.8498 &  50674.61198&  0.196& MDO 2.1m & 5350 & 1.20 & 4.30&  +0.21\\
          &        &  50677.72586&  0.339& MDO 2.1m & 5099 & 0.70 & 3.50&  +0.21\\
          &        &  50737.58366&  0.078& MDO 2.1m & 5786 & 1.10 & 4.70&  +0.15\\
          &        &             &       &          &      &      &     &       \\
\object{SW Vel}    &23.4410 &  49116.49905&  0.348& CTIO 4m  & 5010 & 1.00 & 4.40&  +0.01\\
          &        &             &       &          &      &      &     &       \\
\object{X Pup}    &25.9610 &  50383.94934&  0.308& MDO 2.1m & 5654 & 1.10 & 4.40& --0.00\\
          &        &  51095.95617&  0.734& MDO 2.1m & 4925 & 0.70 & 5.10& --0.08\\
          &        &  50562.60378&  0.190& MDO 2.1m & 6224 & 1.55 & 5.40& --0.06\\
          &        &             &       &          &      &      &     &       \\
\object{T Mon}    &27.0246 &  50379.88342&  0.113& MDO 2.1m & 5811 & 1.40 & 4.90&  +0.12\\
          &        &  50381.97924&  0.191& MDO 2.1m & 5468 & 1.10 & 4.30&  +0.15\\
          &        &  50382.91306&  0.225& MDO 2.1m & 5346 & 1.05 & 4.20&  +0.11\\
          &        &  50383.91603&  0.262& MDO 2.1m & 5238 & 1.00 & 3.90&  +0.12\\
          &        &             &       &          &      &      &     &       \\
\object{SV Vul}    &44.9948 &  48876.62947&  0.669& KPNO 4m  & 4880 & 1.00 & 5.00& --0.03\\
          &        &  48878.76771&  0.717& KPNO 4m  & 4883 & 1.00 & 5.00& --0.04\\
          &        &  49981.67704&  0.237& KPNO 4m  & 5314 & 0.70 & 4.90&  +0.05\\
          &        &  49982.71400&  0.260& KPNO 4m  & 5274 & 0.70 & 4.70&  +0.04\\
          &        &  49983.66534&  0.282& KPNO 4m  & 5209 & 0.70 & 4.70&  +0.02\\
          &        &  49984.66494&  0.304& KPNO 4m  & 5188 & 0.60 & 4.60&  +0.02\\
          &        &  49985.69595&  0.327& KPNO 4m  & 5155 & 0.50 & 4.50& --0.01\\
          &        &  49986.66211&  0.348& KPNO 4m  & 5120 & 0.50 & 4.50&  +0.01\\
          &        &  50379.58714&  0.085& MDO 2.1m & 5805 & 1.20 & 5.80&  +0.08\\
          &        &  50381.55138&  0.129& MDO 2.1m & 5611 & 0.95 & 5.30&  +0.06\\
          &        &  50382.56326&  0.151& MDO 2.1m & 5548 & 0.95 & 5.10&  +0.07\\
          &        &  50383.57136&  0.174& MDO 2.1m & 5432 & 0.80 & 5.00&  +0.02\\
          &        &  50672.75472&  0.604& MDO 2.1m & 4896 & 1.00 & 5.00& --0.03\\
          &        &  50674.77562&  0.649& MDO 2.1m & 4876 & 1.00 & 5.00& --0.02\\
          &        &  50675.75208&  0.671& MDO 2.1m & 4873 & 1.00 & 5.00& --0.03\\
          &        &  50677.79466&  0.716& MDO 2.1m & 4861 & 1.00 & 5.00& --0.04\\
          &        &  50738.69196&  0.070& MDO 2.1m & 5856 & 1.20 & 5.90&  +0.10\\
          &        &  51058.75536&  0.188& MDO 2.1m & 5398 & 0.80 & 5.00&  +0.07\\
          &        &  51094.73986&  0.988& MDO 2.1m & 6110 & 1.60 & 6.80&  +0.04\\
          &        &  51096.69223&  0.031& MDO 2.1m & 5977 & 1.40 & 6.20&  +0.09\\
          &        &  51098.76320&  0.077& MDO 2.1m & 5755 & 1.05 & 5.60&  +0.08\\
          &        &  51473.62697&  0.414& MDO 2.1m & 5005 & 0.50 & 4.80& --0.02\\
          &        &  51473.64877&  0.414& MDO 2.1m & 4995 & 0.50 & 4.80&  +0.00\\
          &        &             &       &          &      &      &     &       \\
\object{S Vul}    &68.4640 &  51055.75037&  0.908& MDO 2.1m & 5881 & 1.10 & 8.70&  +0.02\\
          &        &  51058.73003&  0.951& MDO 2.1m & 5950 & 1.00 & 8.00& --0.05\\
          &        &  51095.66963&  0.491& MDO 2.1m & 5166 & 0.50 & 6.60& --0.05\\
\hline
\end{tabular}
\end{table*}

\section{Methodology}
\subsection{Line equivalent widths}
The line equivalent widths in the Cepheid spectra were measured using the
gaussian approximation, or in some cases, by direct integration. To estimate
the internal accuracy of the measurements we have compared the equivalent
widths of lines present in adjacent overlapping \'echelle orders. In no case
did the difference between independent estimates exceed 10\%. The measured
equivalent widths were analyzed in the LTE approximation using the WIDTH9 code
of Kurucz. Only lines having W$_{\lambda} \le 165$ m\AA~ were used
for abundance determination. The total number of lines used in the analysis
exceeds 41,000.

\subsection{Atmosphere models}
Plane-parallel LTE atmosphere models (ATLAS9) from Kurucz (\cite{kur92}) were
used to determine the elemental abundances. Final models were interpolated (in
T$_{\rm eff}$ and $\log g$) from the solar metallicity grid computed using a
microturbulent velocity of 4 km~s$^{-1}$. The adopted microturbulent velocities
vary from 3 to 8  km~s$^{-1}$ but numerical experiment shows that the derived
abundances are not substantially altered by the mismatch between the model
microturbulence and the value used to computed the abundances. Note that the
ATLAS9 models are in a regime where the convection problem discovered by
Castelli (\cite{cast96}) does not cause a significant problem.

\subsection{Oscillator strengths and damping constants}
The oscillator strengths used in present study were obtained through an
inverted solar analysis (with elemental abundances adopted following Grevesse,
Noels \& Sauval \cite{gns96}). They were derived  using selected unblended solar lines
from the solar spectrum by Kurucz et al. (\cite{kuret84}).
A detailed description and list of the $\log gf$ values can be found in
Kovtyukh \& Andrievsky (\cite{kovan99}). The damping constants for the lines
of interest were taken from the list of B. Bell.  It is well known
that the classical treatment of the van der Waals broadening leads
to broadening coefficients which are too small (see Barklem,
Piskunov, \& O'Mara \cite{bpom00}).

\subsection{Stellar parameters}
The effective temperature for each program star was determined using the
method described in detail in Kovtyukh \& Gorlova (\cite{kovgo00}). That method is based
on the use of relations between effective temperature and the line depth ratios
(each ratio is for the weak lines with different excitation potentials of the
same chemical element).
The advantages of this method are the following: such ratios are sensitive to
the temperature variations, they do not depend upon the abundances and interstellar
reddening. The main source of initial temperatures used for the calibrating
relations was Fry \& Carney (\cite{fc97}). Their data are in  good agreement
with the results obtained by other methods, such as the infrared flux method
(Fernley, Skillen \& Jameson \cite{fsj89}) or detailed analysis of energy distribution
(Evans \& Teays \cite{et96}). Another source was photometry (Kiss \cite{kiss98}).

The number of temperature indicators (ratios) is typically 30. The precision
of the T$_{\rm eff}$ determination is 10-30\,K (standard error of the mean)
from the spectra with S/N greater than 100, and 30-50\,K for S/N less than 100.
Although an internal error of T$_{\rm eff}$ determination appears to be small,
a systematic shift of the zero-point of T$_{\rm eff}$ scale may exist.
Nevertheless, an uncertainty in the zero-point (if it exists) can affect absolute
abundances in each program star, but the slopes of the abundance distributions
should be hardly affected.

The microturbulent velocity and gravity were found using the technique put
forth by Kovtyukh \& Andrievsky (\cite{kovan99}). This method was applied to an
investigation of LMC F supergiants with well known distances, and it produced
much more appropriate gravities for those stars than were previously determined
(see Hill, Andrievsky \& Spite \cite{has95}). The method also allowed to solve
several problems connected with abundances in the Magellanic Cloud supergiants
(for details see Andrievsky et al. \cite{anet01}). Results on T$_{\rm eff}$, $\log g$
and V$_{\rm t}$ determination are gathered in Table 1 (the quoted precision of
the T$_{\rm eff}$ values presented in Table 1 is not representative of the
true precision which is stated above).

Several remarks on the gravity results for our program Cepheids have to be
made. It is expected that the gravities of Cepheids, being averaged over the
pulsational cycle, should correlate with their pulsational periods in the
sense that lower gravities correspond to larger periods. As it was analytically
shown by Gough, Ostriker \& Stobie (\cite{gos65}), the pulsational period P behaves as
P $\sim$ R$^{2}$M$^{-1}$, i.e. $\sim$ g$^{-1}$ (where R, M and g are the radius,
mass and gravity of a Cepheid respectively). A similar relation between
pulsational period and gravity can be also derived, for example, by combining
observational " period-mass" and "period-radius" relations established for
Cepheids by Turner (\cite{turn96}) and Gieren, Fouqu\'e \& G\'omez (\cite{gfg98})
respectively.

As for each star of our sample we have only a limited number of the gravity
estimates, in Fig. 1 we simply plotted individual $\log g$ (gravity in
cm~s$^{-2}$) values for a given Cepheid versus its pulsational period of the
fundamental mode (period is given in days). The general trend can be clearly
traced from this figure. For the long-period Cepheids the scatter in the
gravities derived at the different pulsational phases achieves approximately
1 dex. The instantaneous gravity value, in fact, is a combination of the static
component GM/R$^{2}$ and dynamical term $\gamma$ dV/dt ($\gamma$ is the
projection factor and V is the radial velocity).
This means that observed amlitudes of the gravity variation do not reflect
purely pulsational changes of the Cepheid radius. Nevertheless, there may exist
some additional mechanism artificially "lowering" gravities which are derived
through spectroscopic analysis, and thus increasing an amplitude of the
gravity variation. Such effects as, for example, sphericity of the Cepheid
atmospheres, additional UV flux and connected with it an overionization of
some elements, stellar winds and mass loss, rotation and macroturbulence, may
contribute to some increase of the spectroscopic gravity variation over a
pulsational cycle. It is quite likely that these effects should be more
pronounced in the more luminous (long-period) Cepheids, and they may affect
the abundances resulting from the gravity sensitive ionized species. To
investigate this problem one needs to perform a special detailed multiphase
analysis for Cepheids with various pulsational periods. From our sample of
stars only for TU Cas, U Sgr and SV Vul we have enough data to observe
effects.

In Figs. 2-3 we plotted Ce and Eu abundances together with spectroscopic gravities
versus pulsational phases for the intermediate-period Cepheid U Sgr (P$\approx$ 7
days) and for SV Vul (P$\approx$ 45 days), one of the longest period Cepheid
among our program stars. A scatter of about 0.15 dex is seen for both elements
which are presented in the spectra only by ionized species. Inspecting Fig. 3
one might suspect some small decrease of abundances around phase 0.4
(roughly corresponds to a maximum in SV Vul radius). It can be attributed
to NLTE effects, which should increase in the extended spherical atmosphere
of lower density. More precisely, a small decrease in the abundances may be
caused by additional overionization of the discussed ions having rather low
ionization potentials (about 11 eV). Although the s-process elements in Cepheids
are measured primarily by ionized species, and any errors in the stellar
gravities at some phases propagate directly into the abundance results, we do
not think that this effect may have some radical systematic influence on
abundance results for the s-process elements in our program stars.
The reasons are the following. An indicated decrease in the abundance is rather
small, even for the long-period Cepheid, and practically is not seen for shorter
periods . In fact, a decrease of about 0.15 dex is comparable with errors in the
abundance determination for the elements with small number of lines, like
s-process elements. One can also add that the abundances averaged from different
phases should be sensitive to this effect even to a lesser extent.

\section{Elemental abundances}
Detailed abundance results are presented in the Appendix and Table 2.
The Appendix contains the per species abundances averaged over all individual
spectra (phases) for each star along with the total number of lines and standard
deviation. Table 2 contains the final averaged abundance per element. The
latter abundances were obtained by averaging all lines from all ions at all
phases: that is, for example a sum of all iron line abundances from all phases
and divide by the total number of lines. The iron abundance for individual
phases is presented in Table 1, and the mean iron abundance for each star is
also repeated in Table 3. Note that standard deviation of the abundance of an
element falls within the range 0.05--0.20 dex for all determinations.
Because the number of lines used for some elements is large (especially for
iron), the standard error of the mean abundance is very small.

\begin{table*}
\caption[]{Averaged relative-to-solar elemental abundance for program
Cepheids}
\tiny
\begin{tabular}{lrrrrrrrrrrrrr}
\hline
  Star        &   C  &   O  &  Na  &  Mg  &  Al  &  Si  &   S  &  Ca  &  Sc  &  Ti  &   V  &   Cr &    Mn \\
\hline
 V473 Lyr  (s)&--0.34&--0.24&  0.01&--0.14&--0.06&--0.03&  0.09&--0.10&--0.05&--0.06&--0.04&--0.08&--0.22\\
 SU Cas    (s)&--0.24&--0.02&  0.20&--0.24&  0.05&  0.07&  0.11&--0.01&--0.13&  0.13&  0.08&  0.12&  0.07\\
 EU Tau    (s)&--0.24&--0.05&  0.24&--0.28&--0.01&  0.04&  0.09&--0.05&--0.07&  0.03&--0.05&--0.02&--0.05\\
 IR Cep    (s)&--0.04&    --&  0.28&--0.39&  0.12&  0.17&  0.34&--0.06&--0.03&  0.09&  0.26&--0.12&--0.02\\
 TU Cas       &--0.19&--0.03&  0.15&--0.19&  0.14&  0.10&--0.03&--0.02&--0.19&  0.05&  0.02&  0.02&  0.06\\
 DT Cyg    (s)&--0.12&  0.01&  0.33&  0.04&  0.17&  0.12&  0.15&  0.05&--0.01&  0.21&  0.14&  0.18&  0.19\\
 V526 Mon  (s)&--0.28&--0.52&  0.11&--0.09&    --&--0.01&  0.05&--0.08&--0.20&--0.04&--0.18&--0.02&--0.14\\
 V351 Cep  (s)&--0.19&--0.09&  0.17&--0.30&--0.03&  0.07&  0.26&--0.10&  0.01&  0.08&  0.08&  0.03&--0.03\\
 VX Pup       &    --&    --&  0.08&    --&    --&--0.06&    --&--0.33&--0.13&--0.01&--0.02&--0.16&--0.10\\
 SZ Tau    (s)&--0.20&--0.02&  0.25&    --&  0.14&  0.07&  0.19&  0.02&  0.02&  0.06&  0.10&  0.18&  0.09\\
 V1334 Cyg (s)&--0.30&--0.23&  0.18&--0.31&  0.15&  0.08&  0.11&--0.03&--0.01&  0.02&--0.01&--0.03&  0.03\\
 BG Cru    (s)&--0.18&  0.08&  0.24&    --&  0.29&  0.07&    --&--0.01&--0.33&  0.13&  0.15&--0.04&--0.01\\
 BD Cas    (s)&--0.14&--0.09&--0.03&--0.26&--0.09&  0.03&  0.26&--0.19&--0.23&--0.06&--0.06&--0.14&--0.13\\
 RT Aur       &--0.22&--0.01&  0.29&--0.14&  0.13&  0.12&  0.19&  0.09&  0.05&  0.10&  0.05&  0.08&  0.11\\
 DF Cas       &--0.30&    --&  0.16&--0.33&    --&  0.03&  0.39&--0.18&--0.09&  0.00&--0.01&  0.11&--0.12\\
 SU Cyg       &--0.21&--0.25&  0.23&--0.16&  0.17&  0.04&  0.02&  0.01&--0.11&  0.01&  0.06&  0.02&  0.02\\
 ST Tau       &--0.27&--0.29&  0.23&--0.18&  0.03&  0.03&  0.04&--0.04&--0.11&  0.10&--0.07&--0.04&--0.02\\
 V1726 Cyg (s)&--0.22&    --&  0.29&--0.12&  0.07&  0.11&  0.17&--0.16&--0.05&  0.15&    --&--0.07&  0.00\\
 BQ Ser       &--0.15&--0.13&  0.12&--0.14&  0.14&  0.07&  0.13&--0.05&--0.17&  0.03&--0.01&  0.04&--0.04\\
 Y Lac        &--0.26&--0.37&  0.13&--0.24&  0.13&  0.03&--0.04&--0.05&--0.26&--0.03&  0.02&--0.07&--0.09\\
 T Vul        &--0.26&  0.00&  0.13&--0.31&  0.12&  0.04&  0.12&  0.03&--0.19&  0.00&  0.00&  0.10&  0.01\\
 FF Aql    (s)&--0.31&--0.23&  0.25&--0.24&  0.12&    --&  0.01&--0.03&--0.11&  0.09&  0.14&  0.06&  0.04\\
 CF Cas       &--0.19&  0.06&  0.09&--0.21&  0.10&  0.01&  0.10&--0.01&--0.04&--0.00&--0.06&  0.06&--0.01\\
 BG Lac       &--0.17&  0.10&  0.17&--0.25&  0.08&  0.04&  0.09&--0.01&--0.11&  0.03&--0.05&  0.06&  0.04\\
 Del Cep      &--0.17&  0.01&  0.20&--0.16&  0.16&  0.10&  0.15&  0.01&  0.01&  0.04&  0.09&  0.06&  0.17\\
 V1162 Aql (s)&--0.14&--0.19&  0.13&--0.19&  0.13&  0.06&    --&--0.03&--0.21&--0.03&--0.02&  0.02&--0.01\\
 CV Mon       &--0.25&  0.02&  0.03&--0.32&--0.05&  0.01&  0.08&--0.19&--0.14&  0.10&  0.30&  0.04&--0.05\\
 V Cen        &--0.17&  0.02&  0.01&    --&  0.11&  0.03&  0.14&--0.03&  0.00&  0.09&  0.03&--0.04&--0.11\\
 V924 Cyg (s:)&--0.30&    --&--0.04&--0.38&  0.09&--0.04&  0.05&--0.21&--0.37&--0.21&  0.01&--0.09&  0.10\\
 MY Pup    (s)&--0.36&--0.12&  0.14&--0.36&  0.03&--0.08&--0.14&--0.13&--0.17&--0.09&--0.09&--0.18&--0.24\\
 Y Sgr        &--0.14&--0.06&  0.22&--0.03&  0.27&  0.12&  0.17&  0.04&--0.24&  0.01&  0.07&  0.17&  0.11\\
 EW Sct       &--0.07&--0.04&  0.07&--0.10&  0.15&  0.08&  0.15&--0.01&--0.09&  0.09&  0.08&  0.05&  0.08\\
 FM Aql       &--0.24&--0.19&  0.32&  0.00&  0.33&  0.17&  0.24&  0.17&--0.17&  0.15&  0.10&  0.19&  0.11\\
 TX  Del      &  0.06&  0.16&  0.48&--0.22&    --&    --&    --&  0.16&    --&  0.17&  0.15&    --&  0.38\\
 V367 Sct     &--0.38&    --&  0.21&--0.48&  0.21&  0.05&  0.15&--0.15&  0.07&  0.24&  0.06&--0.05&--0.13\\
 X Vul        &--0.16&  0.03&  0.18&--0.20&  0.17&  0.08&  0.20&--0.04&--0.09&  0.08&  0.04&  0.12&  0.08\\
 AW Per       &--0.23&--0.03&  0.24&--0.27&  0.07&  0.06&  0.18&--0.03&--0.15&  0.01&  0.15&  0.27&  0.10\\
 U Sgr        &--0.16&  0.03&  0.20&--0.17&  0.22&  0.07&  0.15&  0.03&--0.16&  0.06&  0.03&  0.10&  0.06\\
 V496 Aql  (s)&--0.20&--0.15&  0.24&--0.12&  0.10&  0.11&  0.13&--0.03&  0.10&  0.06&  0.06&  0.09&  0.05\\
 Eta Aql      &--0.20&--0.10&  0.19&--0.19&  0.23&  0.12&  0.08&--0.02&--0.18&  0.03&  0.02&  0.22&  0.12\\
 BB Her       &--0.10&  0.04&  0.40&--0.01&  0.23&  0.15&    --&--0.01&  0.07&  0.09&  0.07&  0.10&  0.30\\
 RS Ori       &--0.45&--0.18&  0.06&--0.30&  0.02&  0.02&  0.00&--0.06&--0.19&  0.11&--0.12&--0.09&--0.11\\
 V440 Per  (s)&--0.34&--0.21&  0.05&--0.33&  0.06&  0.00&--0.06&--0.16&--0.14&  0.05&  0.00&--0.06&--0.04\\
 W Sgr        &--0.25&  0.02&  0.18&--0.25&--0.01&  0.04&  0.11&--0.01&--0.12&  0.03&  0.03&  0.03&  0.03\\
 RX Cam       &--0.25&--0.11&  0.18&--0.23&  0.06&  0.05&  0.04&--0.05&  0.04&  0.10&  0.03&  0.08&  0.02\\
 W Gem        &--0.27&--0.12&  0.18&--0.28&  0.10&  0.03&  0.05&--0.08&  0.05&  0.09&  0.00&--0.02&--0.02\\
 U Vul        &--0.18&--0.03&  0.18&--0.16&  0.12&  0.09&  0.17&--0.05&--0.28&  0.05&  0.02&  0.10&  0.01\\
 DL Cas       &--0.31&--0.01&  0.11&  0.11&  0.14&  0.02&  0.21&  0.01&--0.16&  0.03&--0.04&  0.10&  0.07\\
 AC Mon       &--0.42&    --&  0.24&--0.40&    --&--0.02&--0.15&--0.10&--0.07&  0.12&    --&--0.08&--0.19\\
 V636 Cas  (s)&--0.17&--0.09&  0.29&--0.07&  0.09&  0.08&  0.12&  0.15&  0.25&  0.05&  0.05&  0.30&  0.10\\
 S Sge        &--0.12&  0.04&  0.24&--0.22&  0.14&  0.16&  0.17&  0.03&  0.27&  0.12&  0.17&  0.22&  0.19\\
 GQ Ori       &--0.37&--0.12&  0.17&  0.11&  0.04&  0.00&  0.04&--0.25&    --&  0.06&  0.04&  0.01&--0.14\\
 V500 Sco     &--0.20&--0.13&  0.13&--0.23&  0.08&  0.02&  0.06&--0.09&--0.13&--0.06&--0.10&--0.07&--0.03\\
 FN Aql       &--1.31&--0.08&  0.19&--0.21&  0.10&  0.00&--0.02&--0.07&--0.07&  0.04&  0.01&--0.04&--0.12\\
 YZ Sgr       &--0.07&--0.07&  0.31&--0.22&  0.18&  0.12&  0.18&  0.00&    --&  0.03&  0.03&  0.09&  0.07\\
 S Nor        &--0.23&--0.19&  0.27&--0.24&  0.16&  0.05&  0.10&--0.03&  0.01&  0.05&  0.02&--0.08&--0.10\\
 Beta Dor     &--0.31&--0.08&  0.07&--0.30&  0.07&  0.00&--0.04&--0.18&--0.11&  0.01&--0.05&--0.04&  0.03\\
 Zeta Gem     &--0.24&--0.12&  0.25&--0.15&  0.07&  0.04&  0.03&--0.06&  0.13&  0.06&  0.02&  0.00&  0.05\\
 Z Lac        &--0.32&--0.10&  0.23&--0.23&  0.07&  0.05&  0.14&  0.04&--0.18&  0.07&--0.02&  0.05&  0.05\\
 VX Per       &--0.25&--0.15&  0.15&--0.31&  0.03&  0.00&  0.04&--0.12&--0.05&--0.03&--0.11&--0.06&--0.09\\
 V340  Nor(s:)&--0.08&  0.07&  0.29&--0.30&  0.00&  0.03&  0.18&--0.16&--0.12&  0.00&--0.07&  0.01&--0.03\\
 RX Aur       &--0.29&--0.02&  0.18&--0.20&  0.04&  0.05&--0.01&--0.09&--0.31&  0.05&--0.07&  0.04&  0.08\\
 TT Aql       &--0.09&  0.13&  0.28&--0.19&  0.20&  0.11&  0.33&  0.08&  0.12&  0.04&--0.01&  0.09&  0.14\\
 SV Mon       &--0.84&--0.28&  0.28&--0.16&  0.10&  0.00&--0.10&--0.09&--0.30&--0.06&--0.16&--0.07&--0.16\\
 X Cyg        &--0.29&  0.05&  0.26&--0.09&  0.18&  0.13&  0.13&  0.04&    --&  0.21&  0.11&  0.23&  0.11\\
 RW Cam       &--0.14&--0.05&  0.19&--0.23&  0.12&  0.07&  0.18&  0.02&  0.12&  0.04&--0.02&  0.06&  0.05\\
 CD Cyg       &--0.18&--0.11&  0.23&--0.37&  0.19&  0.09&  0.28&  0.08&  0.03&  0.08&  0.06&  0.05&  0.10\\
 Y Oph     (s)&--0.14&--0.01&  0.12&--0.31&  0.14&  0.03&  0.15&--0.15&  0.33&  0.06&  0.05&  0.03&  0.04\\
 SZ Aql       &--0.01&--0.04&  0.28&--0.12&  0.28&  0.17&  0.24&  0.14&  0.11&  0.14&  0.12&  0.19&  0.20\\
 WZ Sgr       &  0.03&  0.13&  0.39&  0.01&  0.28&  0.28&  0.51&  0.19&  0.12&  0.23&  0.17&  0.19&  0.15\\
 SW Vel       &--0.13&  0.18&  0.17&  0.16&  0.04&--0.03&  0.17&--0.16&  0.08&--0.06&--0.14&  0.14&--0.12\\
 X Pup        &--0.29&--0.11&  0.18&    --&  0.10&--0.05&  0.02&  0.06&--0.18&--0.08&--0.16&--0.07&  0.07\\
 T Mon        &--0.27&  0.08&  0.35&    --&  0.10&  0.13&    --&  0.09&    --&  0.06&  0.03&  0.27&  0.14\\
 SV Vul       &  0.02&--0.01&  0.04&--0.10&  0.13&  0.06&  0.16&--0.04&    --&  0.02&--0.08&  0.02&--0.02\\
 S Vul        &--0.34&--0.40&  0.21&    --&  0.22&--0.03&  0.22&--0.04&    --&  0.03&  0.04&  0.10&--0.06\\
\hline
\end{tabular}
\end{table*}

\begin{table*}
{\bf Table 2 (continued)} \\
\tiny
\begin{tabular}{lrrrrrrrrrrrr}
\hline
  Star        &   Fe &   Co &   Ni &   Cu &   Zn &    Y &   Zr &   La &   Ce &   Nd &   Eu &   Gd \\
\hline
 V473 Lyr  (s)&--0.06&--0.13&--0.11&--0.10&  0.12&  0.10&  0.02&  0.20&  0.11&  0.03&--0.02&  0.00\\
 SU Cas    (s)&--0.01&--0.19&  0.00&  0.34&  0.18&  0.19&  0.02&  0.21&--0.04&  0.12&  0.02&--0.20\\
 EU Tau    (s)&--0.06&--0.02&--0.08&  0.21&    --&  0.07&--0.08&  0.16&--0.18&--0.02&  0.01&  0.18\\
 IR Cep    (s)&--0.01&--0.20&--0.07&--0.47&    --&  0.03&  0.16&    --&  0.05&--0.11&    --&    --\\
 TU Cas       &  0.03&--0.08&--0.04&  0.15&  0.46&  0.17&  0.01&  0.24&--0.05&  0.08&  0.11&  0.17\\
 DT Cyg    (s)&  0.11&  0.23&  0.14&  0.42&  0.20&  0.46&--0.06&  0.21&  0.01&  0.25&  0.20&  0.33\\
 V526 Mon  (s)&--0.13&  0.04&--0.07&    --&    --&    --&--0.11&  0.41&    --&  0.25&  0.16&    --\\
 V351 Cep  (s)&  0.03&--0.01&  0.00&--0.05&  0.14&  0.19&--0.01&  0.26&--0.10&  0.12&    --&  0.16\\
 VX Pup       &--0.13&    --&--0.18&  0.25&    --&  0.10&    --&    --&    --&  0.35&--0.15&    --\\
 SZ Tau    (s)&  0.08&--0.01&  0.02&    --&  0.39&  0.17&--0.03&  0.31&  0.06&  0.23&  0.15&    --\\
 V1334 Cyg (s)&--0.04&--0.29&--0.10&  0.39&    --&  0.18&--0.15&  0.21&--0.17&  0.13&  0.13&    --\\
 BG Cru    (s)&--0.02&--0.06&--0.16&--0.68&    --&--0.04&--0.04&    --&  0.28&--0.14&  0.25&    --\\
 BD Cas    (s)&--0.07&  0.09&--0.26&--0.14&    --&  0.01&  0.13&  0.41&  0.19&--0.25&    --&  0.30\\
 RT Aur       &  0.06&--0.09&  0.05&  0.15&  0.24&  0.24&--0.04&  0.14&--0.17&  0.07&  0.01&  0.01\\
 DF Cas       &  0.13&--0.30&  0.04&--0.43&    --&  0.06&  0.22&    --&  0.07&  0.38&  0.28&    --\\
 SU Cyg       &--0.00&--0.01&--0.11&  0.15&    --&  0.16&  0.01&  0.27&--0.19&  0.08&  0.06&  0.10\\
 ST Tau       &--0.05&--0.33&--0.04&  0.19&  0.02&  0.18&--0.13&  0.17&--0.10&  0.14&  0.04&  0.13\\
 V1726 Cyg (s)&--0.02&    --&--0.15&    --&    --&  0.14&    --&    --&    --&  0.24&  0.31&    --\\
 BQ Ser       &--0.04&--0.17&--0.07&  0.08&  0.35&  0.13&--0.10&  0.13&--0.09&  0.22&  0.07&  0.20\\
 Y Lac        &--0.09&    --&--0.15&  0.18&    --&  0.12&--0.24&  0.11&--0.35&  0.16&  0.00&    --\\
 T Vul        &  0.01&  0.01&--0.02&  0.25&  0.30&  0.13&--0.02&  0.20&--0.03&  0.14&  0.08&--0.12\\
 FF Aql    (s)&  0.02&--0.13&  0.01&  0.45&    --&  0.31&--0.11&  0.25&--0.13&  0.08&  0.17&  0.22\\
 CF Cas       &--0.01&--0.15&--0.03&--0.11&  0.25&  0.11&--0.19&  0.14&--0.17&  0.04&  0.06&--0.02\\
 BG Lac       &--0.01&--0.13&--0.03&  0.10&  0.28&  0.14&--0.12&  0.07&--0.17&  0.05&  0.02&  0.18\\
 Del Cep      &  0.06&--0.02&  0.01&  0.55&  0.36&  0.27&--0.14&  0.25&--0.08&  0.17&  0.02&    --\\
 V1162 Aql (s)&  0.01&--0.15&--0.02&  0.28&  0.14&  0.21&--0.21&  0.09&--0.24&  0.02&--0.06&--0.23\\
 CV Mon       &--0.03&    --&--0.08&--0.05&    --&  0.01&  0.00&  0.19&--0.03&  0.29&  0.13&    --\\
 V Cen        &  0.04&--0.21&  0.11&  0.16&  0.37&  0.35&  0.18&  0.26&  0.02&  0.28&  0.20&    --\\
 V924 Cyg (s:)&--0.09&    --&--0.14&  0.12&    --&--0.18&    --&    --&    --&  0.04&    --&    --\\
 MY Pup    (s)&--0.12&--0.10&--0.04&--0.25&--0.09&--0.03&--0.16&  0.17&--0.10&--0.07&--0.04&   -- \\
 Y Sgr        &  0.06&--0.07&  0.03&    --&  0.38&  0.31&--0.08&  0.13&--0.15&  0.10&--0.03&  0.00\\
 EW Sct       &  0.04&--0.10&--0.01&  0.11&  0.34&  0.22&--0.12&  0.29&--0.07&  0.17&  0.06&    --\\
 FM Aql       &  0.08&  0.02&  0.09&  0.26&    --&  0.16&--0.01&  0.23&--0.17&  0.14&  0.09&    --\\
 TX  Del      &  0.24&    --&  0.17&    --&    --&  0.07&    --&  0.13&--0.34&--0.38&    --&    --\\
 V367 Sct     &--0.01&  0.03&--0.01&--0.02&    --&--0.06&--0.15&  0.45&    --&  0.18&  0.36&    --\\
 X Vul        &  0.08&--0.11&  0.07&  0.07&  0.35&  0.23&--0.11&  0.15&--0.15&  0.10&  0.03&  0.04\\
 AW Per       &  0.01&--0.01&  0.04&  0.57&  0.51&  0.11&--0.02&  0.25&--0.11&  0.10&  0.11&--0.12\\
 U Sgr        &  0.04&--0.12&  0.01&  0.02&  0.24&  0.22&--0.11&  0.14&--0.12&  0.05&  0.01&--0.03\\
 V496 Aql  (s)&  0.05&--0.09&  0.03&    --&  0.33&  0.14&--0.10&  0.09&--0.23&  0.02&--0.01&--0.09\\
 Eta Aql      &  0.05&--0.27&  0.04&  0.28&  0.14&  0.23&--0.14&  0.26&--0.19&  0.10&  0.04&--0.05\\
 BB Her       &  0.15&--0.04&  0.15&  0.19&  0.39&  0.32&  0.00&  0.11&--0.17&  0.02&  0.08&    --\\
 RS Ori       &--0.10&--0.01&--0.13&  0.12&  0.21&  0.16&--0.12&  0.18&--0.20&  0.04&  0.00&  0.07\\
 V440 Per  (s)&--0.05&--0.13&--0.05&  0.16&  0.06&  0.26&--0.13&  0.30&--0.10&  0.14&  0.14&  0.12\\
 W Sgr        &--0.01&--0.08&--0.04&  0.21&  0.22&  0.20&--0.11&  0.23&--0.07&  0.06&--0.01&  0.11\\
 RX Cam       &  0.03&--0.14&  0.03&  0.20&  0.14&  0.23&--0.06&  0.27&--0.17&  0.10&  0.09&  0.15\\
 W Gem        &--0.04&--0.21&--0.07&  0.11&  0.16&  0.23&--0.09&  0.28&--0.10&  0.15&  0.10&  0.07\\
 U Vul        &  0.05&--0.09&  0.05&  0.13&  0.31&  0.21&--0.10&  0.13&--0.06&  0.13&  0.02&  0.02\\
 DL Cas       &--0.01&--0.04&  0.00&--0.19&  0.47&  0.21&  0.16&  0.12&  0.09&  0.12&  0.11&--0.05\\
 AC Mon       &--0.07&    --&--0.11&--0.61&    --&  0.05&    --&  0.39&    --&  0.35&  0.28&    --\\
 V636 Cas  (s)&  0.06&--0.09&  0.09&--0.07&  0.30&  0.18&--0.11&  0.16&--0.11&  0.12&  0.02&  0.06\\
 S Sge        &  0.10&  0.00&  0.12&  0.26&  0.39&  0.28&  0.02&  0.29&--0.09&  0.13&  0.15&  0.02\\
 GQ Ori       &--0.03&--0.01&--0.15&    --&    --&  0.29&    --&  0.25&    --&--0.16&  0.09&    --\\
 V500 Sco     &--0.02&--0.18&--0.06&--0.02&  0.21&  0.19&--0.19&  0.18&--0.10&  0.08&  0.03&--0.24\\
 FN Aql       &--0.02&--0.14&--0.05&  0.09&  0.15&  0.16&--0.07&  0.23&--0.09&  0.12&  0.08&  0.12\\
 YZ Sgr       &  0.05&--0.06&  0.03&  0.11&  0.22&  0.35&--0.09&  0.14&--0.10&  0.01&  0.02&--0.01\\
 S Nor        &  0.05&--0.10&--0.07&--0.17&    --&  0.11&  0.14&  0.35&--0.10&  0.00&  0.02&    --\\
 Beta Dor     &--0.01&--0.19&--0.04&--0.45&  0.14&  0.02&  0.03&  0.18&  0.05&--0.05&  0.04&  0.20\\
 Zeta Gem     &  0.04&--0.10&  0.02&  0.03&  0.17&  0.19&--0.07&  0.21&--0.13&  0.08&  0.06&--0.06\\
 Z Lac        &  0.01&--0.13&--0.02&  0.14&  0.25&  0.20&--0.07&  0.26&--0.04&  0.12&  0.06&  0.09\\
 VX Per       &--0.05&--0.20&--0.10&  0.05&  0.18&  0.16&--0.12&  0.21&--0.13&  0.08&  0.03&  0.00\\
 V340  Nor(s:)&  0.00&--0.13&  0.01&--0.06&    --&  0.07&    --&  0.13&--0.25&--0.11&--0.06&  0.12\\
 RX Aur       &--0.07&--0.15&--0.07&  0.28&  0.30&  0.09&--0.11&  0.28&--0.21&  0.08&  0.10&  0.12\\
 TT Aql       &  0.11&--0.07&  0.08&  0.05&  0.41&  0.28&--0.06&  0.20&--0.09&  0.14&  0.05&  0.01\\
 SV Mon       &--0.03&--0.28&--0.12&--0.07&  0.16&  0.26&--0.08&  0.25&--0.14&  0.13&  0.03&    --\\
 X Cyg        &  0.12&  0.06&  0.09&  0.24&  0.35&  0.33&--0.03&  0.29&  0.02&  0.15&  0.14&  0.14\\
 RW Cam       &  0.04&--0.08&  0.05&--0.08&  0.45&  0.17&--0.04&  0.25&--0.11&  0.10&  0.06&  0.02\\
 CD Cyg       &  0.07&--0.03&  0.05&  0.01&  0.32&  0.31&--0.10&  0.23&--0.07&  0.17&  0.08&  0.10\\
 Y Oph     (s)&  0.05&--0.05&  0.03&  0.07&  0.14&  0.33&--0.04&  0.28&--0.07&  0.21&  0.13&  0.08\\
 SZ Aql       &  0.15&--0.02&  0.03&  0.05&  0.42&    --&  0.04&  0.21&--0.12&  0.15&  0.12&  0.12\\
 WZ Sgr       &  0.17&  0.12&  0.21&  0.13&  0.35&  0.30&--0.08&  0.24&--0.08&  0.18&  0.15&  0.08\\
 SW Vel       &  0.01&--0.25&--0.06&--0.41&  0.30&  0.21&    --&  0.23&  0.00&  0.22&  0.16&  0.10\\
 X Pup        &--0.03&--0.25&--0.08&--0.13&  0.26&  0.14&  0.01&  0.27&--0.09&  0.22&  0.09&--0.04\\
 T Mon        &  0.13&--0.03&  0.05&    --&  0.34&    --&  0.04&  0.30&  0.02&  0.22&  0.15&    --\\
 SV Vul       &  0.03&--0.13&  0.00&--0.16&  0.22&  0.26&--0.08&  0.20&--0.13&  0.05&  0.04&  0.03\\
 S Vul        &--0.02&--0.22&  0.02&  0.03&    --&  0.22&--0.06&  0.18&--0.25&  0.15&  0.02&--0.01\\
\hline
\end{tabular}
\end{table*}

Note that the modified method of LTE spectroscopic analysis described in
Kovtyukh \& Andrievsky (\cite{kovan99}) specifies the microturbulent velocity as a
fitting parameter to avoid any systematic trend in the "[Fe/H]$-$EW" relation
based on \ion{Fe}{ii} lines (which are not significantly affected by NLTE
effects unlike \ion{Fe}{i} lines which may be adversely affected). With the
microturbulent velocity obtained in this way, the \ion{Fe}{i} lines
demonstrate a progressively decreasing iron abundance as a function of
increasing equivalent width. Kovtyukh \& Andrievsky (\cite{kovan99}) attribute
this behavior to departures from LTE in \ion{Fe}{i}. To determine the true iron
abundance from \ion{Fe}{i} lines one refers the abundance to the lowest EW,
and it is therefore determined using the [Fe/H]$-$EW relation for these lines
(and this resulting iron abundance from \ion{Fe}{i} lines should be equal to
the mean abundance from \ion{Fe}{ii} provided the surface gravity was properly
chosen).

\ion{Ni}{i} has the second largest number of weak and intermediate strength
lines (after Fe) in almost all program spectra. \ion{Ni}{i} has an atomic
structure similar to that of \ion{Fe}{i}. Therefore, one can suppose that in
many respects it should react to departures from LTE in much the same way as
neutral iron. Thus, to estimate the true nickel content from \ion{Ni}{i} lines
(lines of ionized nickel are not available), we have applied the same method
used for \ion{Fe}{i} lines adopting for the microturbulent velocity the value
determined from \ion{Fe}{ii}; i.e., we have extrapolated the [Ni/H]$-$EW
relation back to the lowest EW and adopted the intercept abundance as indicative
of the true abundance.

Manganese is an important element, but in the available spectral interval it
is represented, as a rule, by only three \ion{Mn}{i} lines with intermediate
equivalent widths. As for nickel, we suppose that the \ion{Mn}{i} ion should be
sensitive to departures from LTE at the same level as \ion{Fe}{i}. Because of
the lack of sufficient numbers of  \ion{Mn}{i} lines it is not possible to proceed
in the same way as with \ion{Fe}{i} lines. Therefore, to estimate the true
manganese content from each spectrum, we used the corresponding dependencies
between iron abundance from \ion{Fe}{i} lines and their equivalent widths, and
corrected the manganese abundance derived from the available \ion{Mn}{i} lines.
The abundance correction for a given equivalent width (EW) of \ion{Mn}{i} line
has been found as $\Delta$[Mn/H] = $\Delta$[Fe/H] = $a\times$EW (where $a$ is
the linear coefficient in the [Fe/H]$-$EW relation for \ion{Fe}{i} lines).

Other ions, as a rule, have lines with smaller equivalent widths (i.e. they
should be less affected by departures from LTE). Abundances of these elements
were found as direct mean values from all appropriate lines. For Ca and Sc
intermediate strength lines are not numerous, while weak lines are often
absent. For these two elements abundance corrections were not determined.
Therefore, their abundances should be interpreted with caution.

\section{Distances}
Galactocentric distances for the program Cepheids were calculated from the
following formula (the distances are given in pc):
\begin{equation}
{\rm  R_{\rm G} = \left[R_{\rm G,\odot}^{2} + (d\cos b)^{2} -
       2R_{G,\odot} d \cos b \cos l\right]^{1/2}}
\end{equation}
where R$_{\rm G,\odot}$ is the galactocentric distance of the Sun, d is
the heliocentric distance of the Cepheid, l is the galactic longitude, and
b is the galactic latitude.  The heliocentric distance d is given by

\begin{equation}
{\rm d = 10^{-0.2 (M_{\rm v}  - <V> - 5 + A_{\rm v})}}
\end{equation}

To estimate the heliocentric distances of program Cepheids we used the
"absolute magnitude - pulsational period" relation of Gieren, Fouqu\'e
\& G\'omez (\cite{gfg98}). E(B-V), $<$B-V$>$, mean visual magnitudes and pulsational
periods are from Fernie et al. (\cite{feret95}), see Table 3.  We use for A$_{\rm v}$
an expression from Laney \& Stobie (\cite{ls93}):

\begin{equation}
{\rm A_{\rm v} = \left[3.07 + 0.28 (B-V)_{0} + 0.04 E(B-V)\right] E(B-V)}
\end{equation}

For s-Cepheids (DCEPS type) the observed periods are those of the first
overtone (see, e.g. Christensen-Dalsgaard \& Petersen \cite{chdp95}). Therefore, for
these stars the corresponding periods of the unexcited fundamental mode were
found using the ratio P$_{1}$/P$_{0} \approx 0.72$, and these periods were
then used to estimate the absolute magnitudes. In the case of V473 Lyr,
the fundamental period P$_{0}$ was found assuming that this star pulsates
in the second overtone (Andrievsky et al. \cite{anet98}), i.e. P$_{1}$/P$_{0} = 0.56$.
For V924 Cyg and V340 Nor, whose association with the group of s-Cepheids is
not certain, we used the observed periods as the fundamental period in order
to estimate M$_{\rm v}$.

The galactocentric distance of the Sun R$_{\rm G,\odot}$ = 7.9 kpc was
adopted from the recent determination by McNamara et al. (\cite{mcnet00}). Estimated
distances and other useful characteristics of our program Cepheids are gathered
in Table 3. Because our spectra were obtained with different spectrographs
having differing resolving powers, and also because for different stars we have
a differing number of spectra (as a rule, one spectrum for Cepheids observed with
6-m telescope), we have assigned for each star a weight in the derivation of the
gradient solution.  We assigned a weight W = 1  to the following stars: those
observed with the 6-m telescope (lower resolution spectra), the two stars
observed by Harris \& Pilachowski (\cite{hp84}), and the stars with one high resolution
spectrum, but a low S/N ratio (VX Pup, CV Mon and MY Pup). For the rest of the
program stars a weight W = 3 was used. The weights are given in the last
column of Table 3. The distribution of the analyzed Cepheids in the galactic
plane is shown in Fig. 4.

\begin{table*}
\caption[]{Some physical and positional characteristics of program Cepheids}
\tiny
\begin{tabular}{ccccccccccc}
\hline
Star &   P, d    &$<$B-V$>$  & E(B-V)  &M$_{\rm v}$& d, pc & l  & b & R$_{\rm G}$, kpc&
$<$[Fe/H]$>$&W\\
\hline
V473 Lyr (s)& 2.6600&0.632&  0.026&--2.47& 517.2 & 60.56&   7.44& 7.66 & --0.06& 3 \\
SU Cas   (s)& 2.7070&0.703&  0.287&--2.49& 322.7 &133.47&   8.52& 8.12 & --0.01& 3 \\
EU Tau   (s)& 2.9200&0.664&  0.172&--2.58&1058.2 &188.80& --5.32& 8.94 & --0.06& 3 \\
IR Cep   (s)& 2.9360&0.870&  0.411&--2.59& 624.4 &103.40&   4.91& 8.07 & --0.01& 1 \\
TU Cas      & 2.1393&0.582&  0.115&--2.21& 821.4 &118.93&--11.40& 8.32 &  +0.03& 3 \\
DT Cyg   (s)& 3.4720&0.538&  0.039&--2.79& 487.4 & 76.55&--10.78& 7.80 &  +0.11& 3 \\
V526 Mon (s)& 3.7150&0.593&  0.093&--2.87&1716.6 &215.13&   1.81& 9.36 & --0.13& 1 \\
V351 Cep (s)& 3.8970&0.940&  0.400&--2.93&1640.5 &105.20& --0.72& 8.48 &  +0.03& 1 \\
VX Pup      & 3.0109&0.610&  0.136&--2.62&1265.5 &237.02& --1.30& 8.65 & --0.13& 1 \\
SZ Tau   (s)& 4.3730&0.844&  0.294&--3.07& 536.5 &179.48&--18.74& 8.41 &  +0.08& 3 \\
V1334 Cyg(s)& 4.6290&0.504&  0.000&--3.14& 633.2 & 83.60& --7.95& 7.85 & --0.04& 3 \\
BG Cru   (s)& 4.6430&0.606&  0.053&--3.14& 491.2 &300.42&   3.35& 7.66 & --0.02& 3 \\
BD Cas (s)  & 3.6510&  -- &  0.734&--3.25&2371.5 &118.00& --0.96& 9.25 & --0.07& 1 \\
RT Aur      & 3.7282&0.595&  0.051&--2.88& 428.2 &183.15&   8.92& 8.32 &  +0.06& 3 \\
DF Cas      & 3.8328&1.181&  0.599&--2.91&2297.9 &136.00&   1.53& 9.68 &  +0.13& 1 \\
SU Cyg      & 3.8455&0.575&  0.096&--2.91& 781.5 & 64.76&   2.50& 7.60 & --0.00& 3\\
ST Tau      & 4.0343&0.847&  0.355&--2.97&1020.7 &193.12& --8.05& 8.89 & --0.05& 3 \\
V1726 Cyg(s)& 5.8830&0.885&  0.312&--3.43&1916.2 & 92.50& --1.61& 8.21 & --0.02& 1 \\
BQ Ser      & 4.2709&1.399&  0.841&--3.04& 911.7 & 35.13&   5.37& 7.18 & --0.04& 3 \\
Y Lac       & 4.3238&0.731&  0.217&--3.05&1996.6 & 98.72& --4.03& 8.43 & --0.09& 3 \\
T Vul       & 4.4355&0.635&  0.064&--3.09& 532.7 & 72.13&--10.15& 7.76 &  +0.01& 3 \\
FF Aql   (s)& 6.2100&0.756&  0.224&--3.49& 424.4 & 49.20&   6.36& 7.63 &  +0.02& 3 \\
CF Cas      & 4.8752&1.174&  0.566&--3.20&3145.2 &116.58& --0.99& 9.72 & --0.01& 3 \\
TV Cam      & 5.2950&1.198&  0.644&--3.30&3739.1 &145.02&   6.15&11.15 & --0.06& 1 \\
BG Lac      & 5.3319&0.949&  0.336&--3.31&1656.6 & 92.97& --9.26& 8.15 & --0.01& 3 \\
Del Cep     & 5.3663&0.657&  0.092&--3.31& 247.9 &105.19&   0.53& 7.97 &  +0.06& 3 \\
V1162 Aql(s)& 7.4670&0.900&  0.205&--3.71&1470.2 & 29.40&--18.60& 6.72 &  +0.01& 3 \\
CV Mon      & 5.3789&1.297&  0.714&--3.32&1809.1 &208.57& --1.79& 9.53 & --0.03& 1 \\
V Cen       & 5.4939&0.875&  0.289&--3.34& 702.3 &316.40&   3.31& 7.41 &  +0.04& 3 \\
V924 Cyg    & 5.5710&0.847&  0.258&--3.36&4428.4 & 66.90&   5.33& 7.38 & --0.09& 1 \\
MY Pup   (s)& 7.9100&0.631&  0.064&--3.78& 708.4 &261.31&--12.86& 8.03 & --0.12& 1 \\
Y Sgr       & 5.7734&0.856&  0.205&--3.40& 496.1 & 12.79& --2.13& 7.42 &  +0.06& 3 \\
EW Sct      & 5.8233&1.725&  1.128&--3.41& 345.0 & 25.34& --0.09& 7.59 &  +0.04& 3 \\
FM Aql      & 6.1142&1.277&  0.646&--3.47& 842.3 & 44.34&   0.89& 7.32 &  +0.08& 3 \\
TX  Del     & 6.1660&0.766&  0.132&--3.48&2782.5 & 50.96&--24.26& 6.60 &  +0.24& 3 \\
V367 Sct    & 6.2931&1.769&  1.130&--3.51&1887.7 & 21.63& --0.83& 6.18 & --0.01& 3 \\
X Vul       & 6.3195&1.389&  0.848&--3.51& 831.6 & 63.86& --1.28& 7.57 &  +0.08& 3 \\
AW Per      & 6.4636&1.055&  0.534&--3.54& 724.9 &166.62& --5.39& 8.60 &  +0.01& 3 \\
U Sgr       & 6.7452&1.087&  0.403&--3.59& 620.5 & 13.71& --4.46& 7.30 &  +0.04& 3 \\
V496 Aql (s)& 9.4540&1.146&  0.413&--4.00&1195.1 & 28.20& --7.13& 6.88 &  +0.05& 3 \\
Eta Aql     & 7.1767&0.789&  0.149&--3.66& 260.1 & 40.94&--13.07& 7.71 &  +0.05& 3 \\
BB Her      & 7.5080&1.100&  0.414&--3.72&3091.7 & 43.30&   6.81& 6.04 &  +0.15& 3 \\
RS Ori      & 7.5669&0.945&  0.389&--3.73&1498.9 &196.58&   0.35& 9.35 & --0.10& 3 \\
V440 Per (s)&10.5140&0.873&  0.273&--4.12& 801.1 &135.87& --5.17& 8.49 & --0.05& 3 \\
W Sgr       & 7.5949&0.746&  0.111&--3.73& 405.4 &  1.58& --3.98& 7.50 & --0.01& 3 \\
RX Cam      & 7.9120&1.193&  0.569&--3.78& 833.4 &145.90&   4.70& 8.60 &  +0.03& 3 \\
W Gem       & 7.9138&0.889&  0.283&--3.78& 916.9 &197.43&   3.38& 8.78 & --0.04& 3 \\
U Vul       & 7.9906&1.275&  0.654&--3.79& 570.9 & 56.07& --0.29& 7.60 &  +0.05& 3 \\
DL Cas      & 8.0007&1.154&  0.533&--3.79&1602.1 &120.27& --2.55& 8.82 & --0.01& 3 \\
AC Mon      & 8.0143&1.165&  0.508&--3.80&2754.8 &221.80& --1.86&10.12 & --0.07& 1 \\
V636 Cas (s)&11.6350&1.391&  0.786&--4.25& 592.2 &127.50&   1.09& 8.27 &  +0.06& 3 \\
S Sge       & 8.3821&0.805&  0.127&--3.85& 648.1 & 55.17& --6.12& 7.55 &  +0.10& 3 \\
GQ Ori      & 8.6161&0.976&  0.279&--3.88&2437.6 &199.77& --4.42&10.22 & --0.03& 1 \\
V500 Sco    & 9.3168&1.276&  0.599&--3.98&1406.1 &359.02& --1.35& 6.49 & --0.02& 3 \\
FN Aql      & 9.4816&1.214&  0.510&--4.00&1383.2 & 38.54& --3.11& 6.87 & --0.02& 3 \\
YZ Sgr      & 9.5536&1.032&  0.292&--4.01&1205.4 & 17.75& --7.12& 6.77 &  +0.05& 3 \\
S Nor       & 9.7542&0.941&  0.215&--4.03& 879.6 &327.80& --5.39& 7.17 &  +0.05& 3 \\
Beta Dor    & 9.8424&0.807&  0.044&--4.04& 335.8 &271.73&--32.78& 7.90 & --0.01& 3 \\
Zeta Gem    &10.1507&0.798&  0.018&--4.08& 387.2 &195.75&  11.90& 8.27 &  +0.04& 3 \\
Z Lac       &10.8856&1.095&  0.404&--4.17&1782.4 &105.76& --1.62& 8.56 &  +0.01& 3 \\
VX Per      &10.8890&1.158&  0.515&--4.17&2283.5 &132.80& --2.96& 9.60 & --0.05& 3 \\
V340 Nor(s:)&11.2870&1.149&  0.332&--4.21&1976.1 &329.80& --2.23& 6.27 &  +0.00& 3 \\
RX Aur      &11.6235&1.009&  0.276&--4.24&1579.0 &165.77& --1.28& 9.44 & --0.07& 3 \\
TT Aql      &13.7547&1.292&  0.495&--4.45& 976.1 & 36.00& --3.14& 7.13 &  +0.11& 3 \\
SV Mon      &15.2328&1.048&  0.249&--4.57&2472.3 &203.74& --3.67&10.21 & --0.03& 3 \\
X Cyg       &16.3863&1.130&  0.288&--4.66&1043.5 & 76.87& --4.26& 7.73 &  +0.12& 3 \\
RW Cam      &16.4148&1.351&  0.649&--4.66&1748.4 &144.85&   3.80& 9.38 &  +0.04& 3 \\
CD Cyg      &17.0740&1.266&  0.514&--4.71&2462.1 & 71.07&   1.43& 7.47 &  +0.07& 3 \\
Y Oph    (s)&23.7880&1.377&  0.655&--5.11& 664.9 & 20.60&  10.12& 7.29 &  +0.05& 3 \\
SZ Aql      &17.1408&1.389&  0.641&--4.71&1731.0 & 35.60& --2.34& 6.57 &  +0.15& 3 \\
YZ Aur      &18.1932&1.375&  0.565&--4.78&4444.7 &167.28&   0.94&12.27 & --0.05& 1 \\
WZ Sgr      &21.8498&1.392&  0.467&--5.00&1967.5 & 12.11& --1.32& 5.99 &  +0.17& 3 \\
SW Vel      &23.4410&1.162&  0.349&--5.09&2572.4 &266.20& --3.00& 8.47 &  +0.01& 3 \\
X Pup       &25.9610&1.127&  0.443&--5.21&2776.6 &236.14& --0.78& 9.72 & --0.03& 3 \\
T Mon       &27.0246&1.166&  0.209&--5.26&1369.7 &203.63& --2.55& 9.17 &  +0.13& 3 \\
SV Vul      &44.9948&1.442&  0.570&--5.87&1729.5 & 63.95&   0.32& 7.31 &  +0.03& 3 \\
S Vul       &68.4640&1.892&  0.827&--6.38&3199.7 & 63.45&   0.83& 7.07 & --0.02& 3 \\
\hline
\end{tabular}
\end{table*}

\section{Results and discussion}
\subsection{The radial distribution of elemental abundances:
general picture and remarks on some elements}
Using our calculated galactocentric distances and average abundances we
can determine the galactic metallicity gradient from a number of species.
Plots for several chemical elements and results of a linear fit are given in
Fig. 5 (iron) and Figs. 6-9 (other elements). Note that, in the plots for Si
and Cr, TX Del is not included. This star shows rather strong excess in the
abundances of these elements which could be connected with its peculiar nature
(in Harris \& Welch \cite{hw89} TX Del is reported as a spectroscopic binary. It has
also been labeled a Type II Cepheid at times). In the plot for carbon we did
not include the data for FN Aql and SV Mon, both of which have an extremely
low carbon abundances. These two unusual Cepheids will be discussed in detail
in a separate paper.

The information in the plots and also in Fig. 10 enables one to put together
several important
conclusions.
Most radial distributions of the elements studied indicate a negative gradient
ranging from about $-0.02$ dex~kpc$^{-1}$ to $-0.06$ dex~kpc$^{-1}$, with an
average of $-0.03$ dex~kpc$^{-1}$ for the elements in Figs. 5-8.
The most reliable value comes from iron (typically the number of iron lines
for each star is about 200-300).
The gradient in iron is $-0.029$ dex~kpc$^{-1}$, which is close to the typical
gradient value produced by other iron-group elements. Examination of Fig. 5
might lead one to suspect that the iron gradient is being controlled by the
cluster of stars at R $\approx 6.5$ with [Fe/H] $\approx 0.2$. If one deletes
these stars from the solution the gradient falls to approximately $-0.02$
dex~kpc$^{-1}$. This latter value differs from the value determined using
all the data by about twice the formal uncertainty in either slope. However,
we do not favour the neglect these points as there is no reason to suspect
these abundances relative to the bulk of the objects. Indeed, in a subsequent
paper, we shall present results for Cepheids which lie closer to the galactic
center and which have abundances above those of this study, which may imply
a steepening of the gradient towards the galactic center.

Unweighted iron abundances give a gradient of $-0.028$ dex~kpc$^{-1}$.
Both weighted and unweighted iron gradients are not significantly changed
if we remove two Cepheids at galactocentric distances greater than 11 kpc
(gradient is $-0.031$ dex~kpc$^{-1}$).
Thus, the average slope of about $-0.03$ dex~kpc$^{-1}$ probably
applies to the range $6 \le$ R$_{\rm G}$ (kpc) $\le 10$. Notice
that in all cases the correlation coefficient is relatively low, r
$\approx 0.47$.

Carbon shows a surprisingly clear dependence upon galactocentric distance
(Fig. 6a): the slope of the relation is among the largest from examined
elements. We have included in the present study elements such as carbon
and sodium, although the gradients based on their abundances determined
from Cepheids may not be conclusive. In fact, it is quite likely that the
surface abundances of these elements have been altered in these intermediate
mass stars during their evolution from the main sequence to the Cepheid stage.
For example, the surface abundance of carbon should be decreased after the
global mixing which brings the CNO-processed material into the stellar atmosphere
(turbulent diffusion in the progenitor B main sequence star, or the first
dredge-up in the red giant phase). Some decrease in the surface abundance of
oxygen is also expected for supergiant stars, but at a significantly lower
level than for carbon (Schaller et al. \cite{schallet92}).

It is also well known that galactic supergiants (Cepheids, in particular) show
an increased sodium abundance which is usually interpreted as a result of
dredge-up of material processed in the Ne--Na cycle (and therefore enriched
in sodium) to the stellar surface (Sasselov \cite{sass86}, Luck \cite{luck94},
Denissenkov \cite{denis93a}, \cite{denis93b}, \cite{denis94}).
Such a contamination of the Cepheids' atmospheres with additional
sodium may result in a bias of the [Na/H] gradient value derived from
Cepheids in relation to the true gradient. It can be seen that our results
in Fig. 6a,c are consistent with these considerations on C and Na,
respectively. It is not clear how these effects should affect stars
at different galactocentric distances (with different metallicities),
but it is likely that they contribute to increase the dispersion in the
abundances, thus producing a flatter gradient.

There are some indications (Andrievsky \& Kovtyukh \cite{ankov96}) that surface
Mg and Al abundances in yellow supergiants can be altered to some extent due
to mixing of the material processed in the Mg--Al cycle with atmospheric gas.
This supposition seems to gain some additional support from our present data
(see Fig.11) where one can see that the Mg and Al abundances are correlated.

As surface abundance modifications depend upon the number of visits to the red
giant region (i.e. the number of dredge-up events) as well as other factors
(pre dredge-up events, depth of mixing events, mass), it is possible that the
program Cepheids could show differential evolutionary effects in their abundances.
Because of the high probability of such effects impacting the observed carbon
and sodium (and perhaps, oxygen, magnesium, and aluminum) abundances in these
Cepheids, we recommend that our gradient values for carbon and sodium to be viewed
with extreme caution, while the gradients of oxygen, magnesium and aluminum
abundances could be used, but also with some caution.

The difference in metallicity between the stars of our sample (say, at 6 kpc
and 10 kpc) is about 0.25 dex. This is a rather small value to detect/investigate
the so-called "odd-even" effect, that is the metallicity dependent yield for
some elements which should be imprinted on the trends of abundance ratios for
[El$_{\rm odd}$/El$_{\rm even}$] versus galactocentric distance, see for details
Hou, Prantzos \& Boissier (\cite{hpb00}). Such elements as, for example, aluminum,
scandium, vanadium and manganese should show  progressively decreasing abundances
with overall metal decrease. This should manifest itself as a gradient in
[El$_{\rm odd}$/Fe]. We have plotted the abundance ratios for some "odd"
elements (normalized to iron abundance) versus R$_{\rm G}$ in Fig.12. As one
can see, none of the abundance ratios plotted versus galactocentric distance
shows a clear dependence upon R$_{\rm G}$. This could mean that the "odd-even"
effect may be overestimated if only the yields from massive stars are taken
into account ignoring other possible sources, or that the effect is not
sufficiently large to be seen over the current distance and metallicity
baseline.

%

\subsection{Metallicity dispersion and the metallicity in the solar
vicinity}

There is a spread in the metallicity at each given galactocentric distance
(larger than the standard error of the abundance analysis) which is most likely
connected with local inhomogeneities in the galactic disc. As an
example, in Fig. 13, we show the derived iron abundance vs. galactic
longitude for the stars of our sample (a few Cepheids with
heliocentric distances large than 3000 pc were excluded). The
distribution gives only a small hint about a local increase of the
metallicity in the solar vicinity towards the direction l$\approx
30^{\degr}$ and $150^{\degr}$.

It is important to note that at the solar galactocentric distance those
elements, whose abundance is not supposed to be changed in supergiants during
their evolution, show on average the solar abundance in Cepheids.
Relative to the solar region, the stars within our sample which are within
500 pc of the Sun have a mean [Fe/H] of $\approx +0.01$ (n = 14, $\sigma = 0.06$).
If we consider all program stars at a galactocentric radius of 7.4--8.4 kpc,
i.e. those in a 1 kpc wide annulus centered at the solar radius, we find a
mean [Fe/H] of approximately +0.03 (n = 29, $\sigma = 0.05$).

This result again stresses the importance of the problem connected with
subsolar metallicities reported for the hot stars from the solar vicinity
(see, e.g. Gies \& Lambert \cite{gila92}, Cunha \& Lambert \cite{cunla94}, Kilian
\cite{kil92}, Kilian, Montenbruck \& Nissen \cite{kmn94}, Daflon, Cunha \& Becker
\cite{dafcube99}, Andrievsky et al. \cite{anet99}). This also follows from
the plots provided by Gummersbach et al. (\cite{guet98}) for several elements.

This problem was discussed, for instance, by Luck et al. (\cite{lucket00}).
The authors compared the elemental abundances of B stars from the open cluster
M25 with those of the Cepheid U Sgr and two cool supergiants which are also members
of the cluster, and found disagreement in the abundances of the B stars and
supergiants; e.g., while the supergiants of M25 show nearly solar abundances,
the sample of B stars demonstrate a variety of patterns from under- to
over-abundances. This should not be observed if we assume that all stars in
the cluster were born from the same parental nebula. Obviously, the problem
of some disagreement between abundance results from young supergiants and
main-sequence stars requires further investigation.

\subsection{Flattening of the elemental distribution in the solar
neighbourhood}

All previous studies of the radial abundance distribution in the galactic disc
have considered only chemical elements from carbon to iron, and all derived
gradients have shown a progressive decrease in abundance with increasing
galactocentric distance. For the elements from carbon to yttrium in this study
our gradient values also have negative signs, while for the heavier species
(from zirconium to gadolinium) we obtained (within the error bars) near-to-zero
gradients (see Fig. 10). Two obvious features which are inherent to derived
C-Gd abundance distributions have to be interpreted: a rather flat character
of the distribution for light/iron-group elements, and an apparent absence of
a clear gradient for heavy species.

The flattening of the abundance distribution can be caused by radial flows
in the disc which may lead to a homogenization of ISM. Among the possible
sources forcing gas of ISM to flow in the radial direction, and therefore
producing a net mixing effect there could be a gas viscosity in the disc,
 gas infall from the halo, gravitational interaction between gas and spiral
waves or a central bar (see e.g., Lacey \& Fall \cite{lf85} and Portinari
\& Chiosi \cite{pch00}).

The mechanism of the angular momentum re-distribution in the disc based
on the gas infall from the halo is dependent upon the infall rate, and
therefore it should have been important at the earlier stages of the Galaxy
evolution, while other sources of the radial flows should effectively
operate at present.

Gravitational interaction between the gas and density waves produces the
radial flows with velocity (Lacey \& Fall \cite{lf85}):
\begin{equation}
           v_{\rm r} \sim (\Omega_{\rm p} - \Omega_{\rm c})^{-1},
\end{equation}
where $\Omega_{\rm p}$ and $\Omega_{\rm c}$ are the angular velocities
of the spiral wave and the disc rotation respectively. According to Amaral
\& L\'epine (\cite{amle97}) and Mishurov et al. (\cite{miet97}) among others,
based on several different arguments, the galactic co-rotation resonance is
located close to the solar galactic orbit. The co-rotation radius is the radius
at which the galactic rotation velocity coincides with the rotation velocity of
the spiral pattern. Together with Eq. (4) this means that, inside the co-rotation
circle, gas flows towards the galactic center ($\Omega_{\rm c} > \Omega_{\rm p}$
and v$_{\rm r} < 0$), while outside it flows outwards. This mechanism can
produce some "cleaning" effect in the solar vicinity, and thus can lead to some
flattening of the abundance distribution. In addition, it could explain
the similarity in the solar abundances and mean abundances in the five billion
years younger Cepheids located at the solar galactocentric annulus (see Fig. 5),
although one might expect that the Cepheids from this region should be more
abundant in metals than our Sun.

There is a clear evidence that the bars of spiral galaxies have also a great
impact on chemical homogenization in the discs (Edmunds \& Roy \cite{er93}; Martin
\& Roy \cite{mr94}, Gadotti \& Dos Anjos \cite{gda01}). It has been shown that
a flatter abundance gradient is inherent to galaxies which have a bar structure.
This could imply that a rotating bar is capable of producing significant
homogenization of the interstellar medium, while such homogenization is not
efficient in unbarred spiral galaxies.

The direct detection of a bar at the center of our galaxy using $COBE$ maps
was reported by Blitz \& Spergel (\cite{bs91}). Kuijken (\cite{kui96}), Gerhard
(\cite{ger96}), Gerhard, Binney \& Zhao (\cite{geret98}), Raboud et al.
(\cite{rabet98}) also suggest that the Milky Way is a barred galaxy. The most
recent evidence for a long thin galactic bar was reported by L\'opez-Corredoira
et al. (\cite{lcet01}) from the DENIS survey. These authors conclude that our
Galaxy is a typical barred spiral. If so, then the Milky Way should obey the
relation between the slope of metallicity distribution and the bar strength
(specifically, the axial ratio), which is based on the data obtained from other
galaxies.

According the above mentioned authors
the galactic bar is triaxial and has an axial ratio (b/a) of about $1/3-1/2$
(see also Fux \cite{fux97}, \cite{fux99}). With such axial ratio an ellipticity
E$_{\rm B}$ = 10 (1-b/a) $\approx 5-7$. L\'epine \& Leroy (\cite{ll00}) presented a model
which reproduces a near-infrared brightness distribution in the Galaxy. Their
estimate of the galactic bar characteristics supposes that the total length of
the bar should be about 4.6 kpc, while its width about 0.5 kpc. In this case
an ellipticity could be even larger than 7. For such ellipticities the
observational calibration of Martin \& Roy (\cite{mr94}) for barred galaxies
predicts a metallicity slope of about $-0.03$ to $\approx 0$ dex~kpc$^{-1}$
for oxygen. Our results on abundance gradients in the solar neighbourhood for
iron-group elements and light species (such as Si, Ca, and even oxygen) appear
to be in good agreement with expected gradient value which is estimated for
the galactic disc solely from bar characteristics.

Martinet \& Friedli (\cite{mf97}) investigated secular chemical evolution in
barred systems and found that a strong bar is capable of producing significant
flattening of the initial gradient across the disc. Using numerical results
of that paper one can trace the (O/H) abundance evolution in barred systems.
With our abundance gradients for such elements as oxygen, silicon, calcium
and iron-group elements one can conclude that an expected age of the galactic
bar is approximately 1 Gyr, or less. Another important result obtained by
Martinet \& Friedli (\cite{mf97}) is that the bar of such an age should produce not
only significant flattening across almost the whole disc, but also
steepening of the abundance distribution in the inner parts (our observational
results for this region will be discussed in the next paper from this series).

An additional mechanism which may cause some local flattening (or even a shallow
local minimum in the elemental abundances) should operate near the galactocentric
solar radius where the relative rotational velocity of the disc and spiral
pattern is small. The shocks that arise when the gas orbiting in the
disc penetrates the spiral potential perturbation, and which are responsible
for triggering star formation in spiral arms, pass through a minimum strength
at this galactic radius, due to almost zero relative velocity.
Furthermore, simulations performed by L\'epine, Mishurov \& Dedikov (\cite{lmd01})
show that there is also a gas depletion at the co-rotation radius. Both reasons
point towards a minimum of star formation rate at the co-rotation radius.
This lower star formation rate manifests itself in the models as a minimum in
elemental abundances. One can expect that after a few billion years, a galactic
radius with minimum star formation rate should correspond to a local minimum
in metallicity. The flat local minimum in metal abundance should be observable,
unless the mechanisms that produce radial transport or radial mixing of the
gas in the disc are important, or if the co-rotation radius varied appreciably
in a few billion years. Note that the star-formation rate also depends on the
gas density, which decreases towards large galactic radii. The combined effect
of gas density and co-rotation could produce a slightly displaced minimum.

At first glance, the abundance data presented in Figs. 5-9 show
little indication of a local abundance minimum (or discontinuity) at the
solar galactocentric radius. Nevertheless, the parabolic fit of the iron
abundance distribution rather well represents observed data, and shows that
a small increase in the metallicity at galactic radii larger than the
co-rotation radius may not be excluded (Fig. 14).

Comparing gradients from iron-group elements (small and negative) with
those from the heaviest species (near to zero) one could propose the following
preliminary explanation of the observed difference. The known contributors of
the O-to-Fe-peak nuclei to ISM are massive stars exploding as SNe II
(short-lived) and SNe I (long-lived), while s-process elements (past iron-peak)
are created only in the low-mass AGB stars (1-4 M$_{\odot}$, Travaglio et al.
\cite{travet99}). The extremely flat distribution in the disc seen for s-process elements
implies that there should exist some mechanism(s) effectively mixing ISM at
time-scales less than the life times of the stars with masses 1-4 M$_{\odot}$
($\tau \approx 0.3-10$ Gyr). At the same time such a mechanism may not be able
to completely erase the O-Fe gradients related to the ISM, and imprinted on the
young stars. If the characteristic time of the mixing (even being possibly
comparable to the SNe I life time) exceeds a nuclear evolution of the SNe II
O-Fe contributors, then these are the high-mass stars that could be responsible
for the resulting small negative gradients from O-Fe elements in the disc.

If one adopts the velocity of the radial flows, say, 4 km~s$^{-1}$ (see
discussion in Lacey \& Fall \cite{lf85} and Stark \& Brand \cite{stbr89}), then
the necessary time to mix the gas within about 4 kpc (baseline covered by our data)
should be likely less than 1 Gyr, that is below the life-time intervals for AGB
progenitors with 1-2 M$_{\odot}$.
However, this $ad~hoc$ supposition meets a problem with the observed Eu gradient.
This element is believed to be produced mainly through the r-process in
lower-mass SNe II (e.g., Travaglio et al. \cite{travet99}), and therefore should
probably behave similar to, for example, iron, but its radial abundance distribution
appears to be quite similar to that of the s-process elements, like Zr, La, Ce,
Nd (see Fig. 10).

\begin{acknowledgements}
SMA would like to express his gratitude to FAPESP for the visiting
professor fellowship (No. 2000/06587-3) and to Instituto Astron\^ {o}mico
e Geof\' \i sico, Universidade de S\~{a}o Paulo for providing facility
support during a productive stay in Brazil.

The authors thank Drs. A. Fry and B.W. Carney for the CCD spectra of some
Cepheids, Drs. H.C. Harris and C.A. Pilachowski for providing the
plate material on two distant Cepheids TV Cam and YZ Aur, Dr. G.A. Galazutdinov
for the spectra of V351 Cep, BD Cas and TX Del, Dr. I.A. Usenko,
Mrs. L.Yu. Kostynchuk and Mr. Yu.V.Beletsky for the help with data reduction.
The authors are also thankful to Dr. Yu.N. Mishurov for discussion and
Drs. N. Prantzos, J.L. Hou and C. Bertout for several comments.

We are indebted to the referee, Dr. B.W. Carney, for a detailed reading of the
paper, and for his many valuable suggestions and comments which improved the
first version.
\end{acknowledgements}

{}

\newpage
\begin{figure*}
\resizebox{\hsize}{!}{\includegraphics{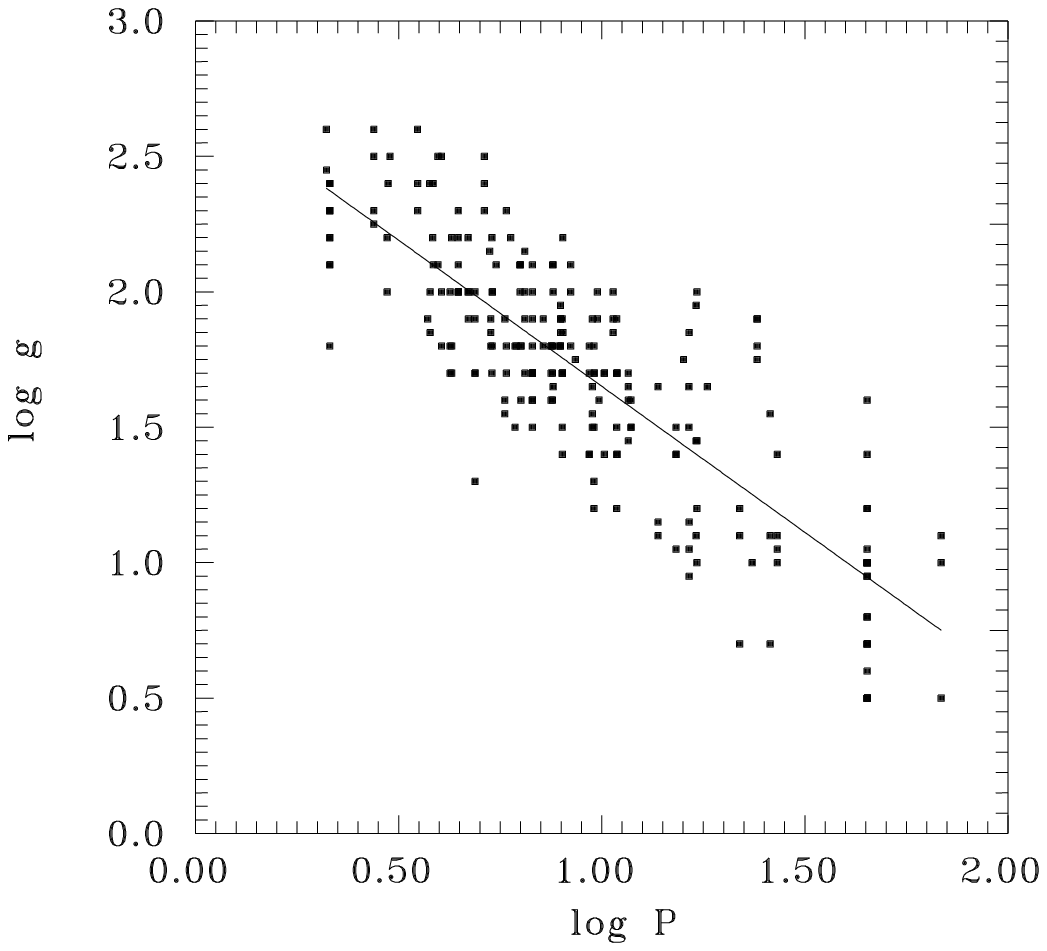}}
\caption[]{Program Cepheid gravities vs. their pulsational periods}
\end{figure*}
\newpage
\begin{figure*}
\resizebox{\hsize}{!}{\includegraphics{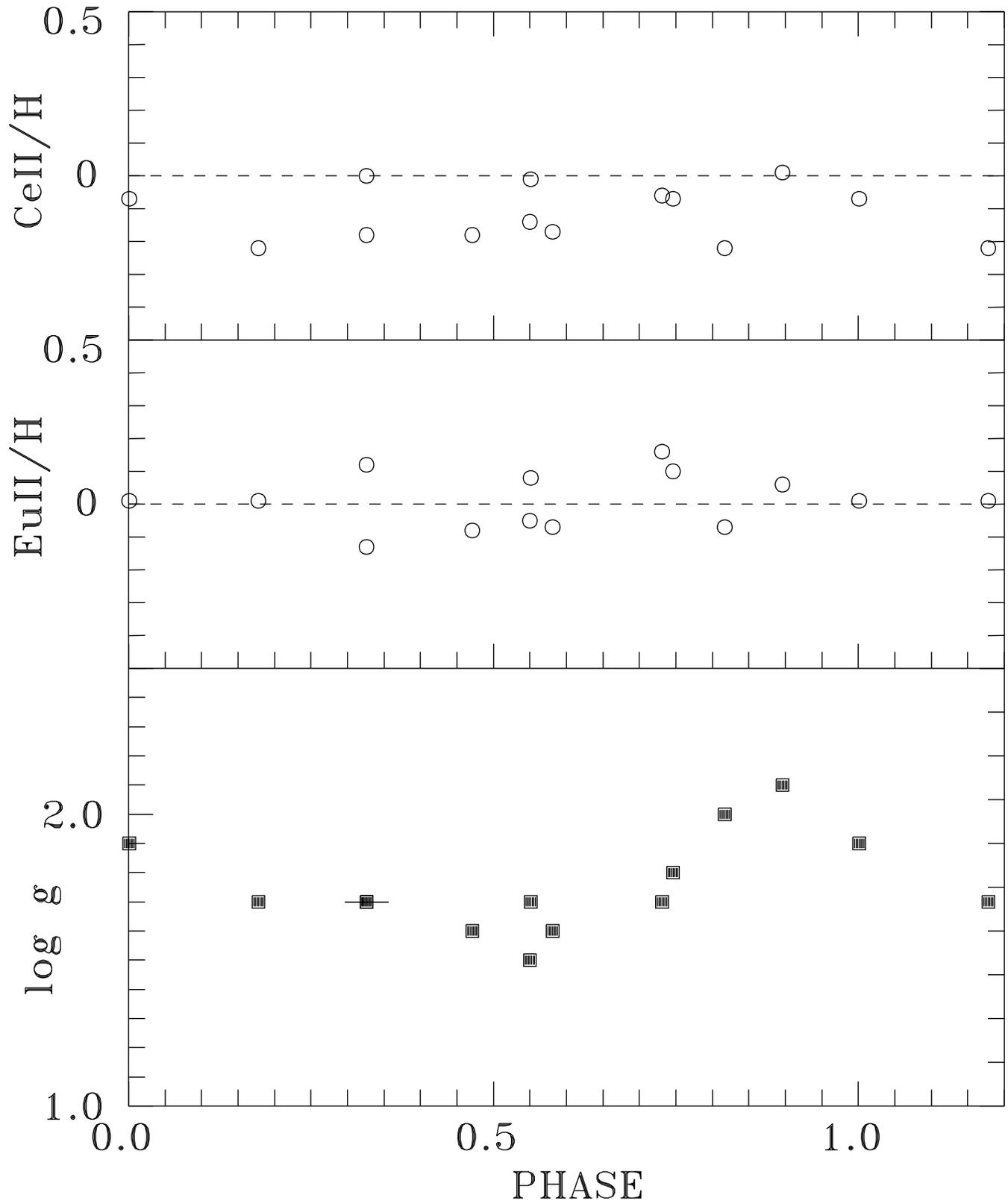}}
\caption[]{Relative-to-solar Ce and Eu abundance and spectroscopic gravity
of U Sgr vs. its pulsational phases}
\end{figure*}
\newpage
\begin{figure*}
\resizebox{\hsize}{!}{\includegraphics{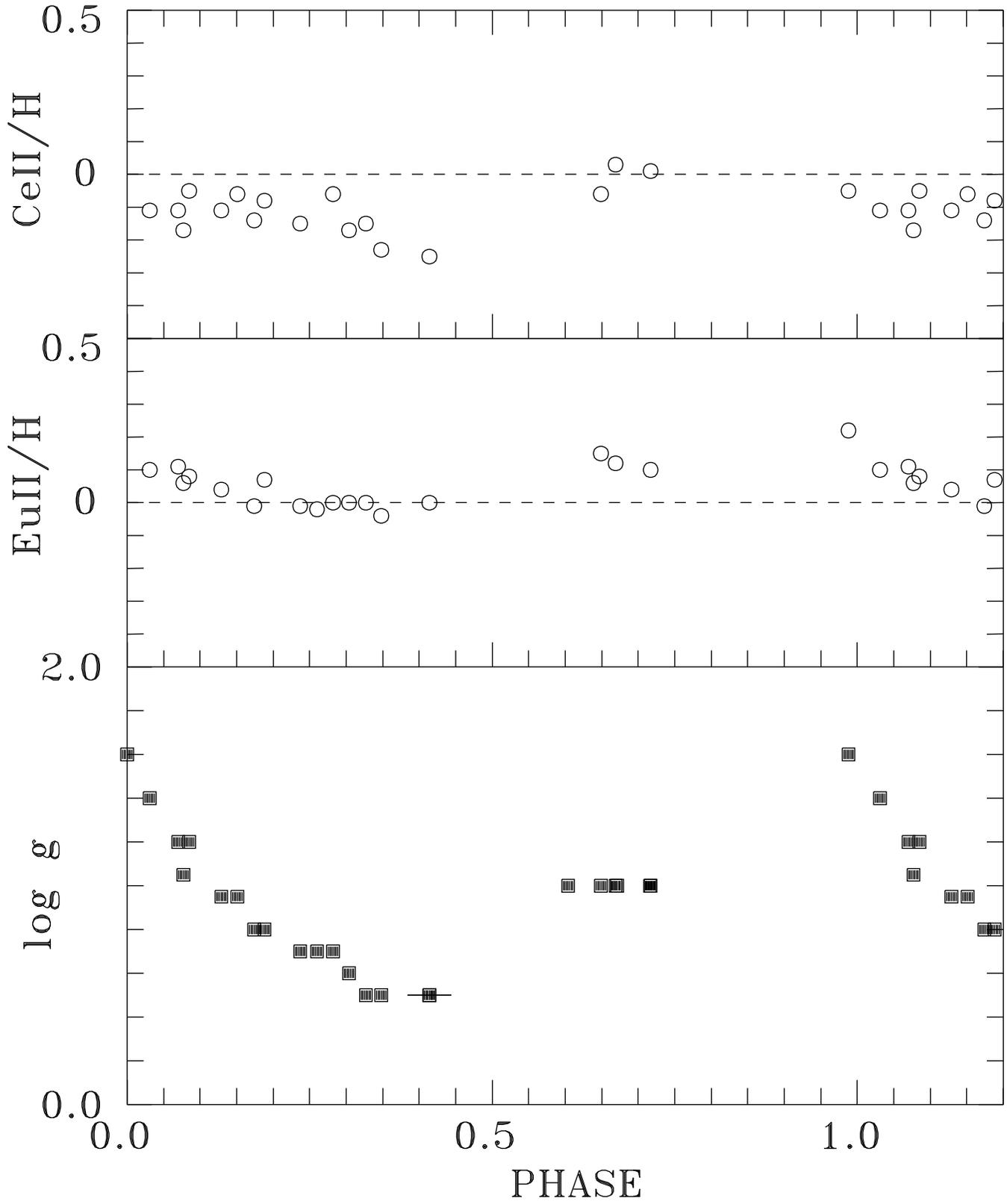}}
\caption[]{Same as Fig. 2 but for SV Vul}
\end{figure*}
\newpage
\begin{figure*}
\resizebox{\hsize}{!}{\includegraphics{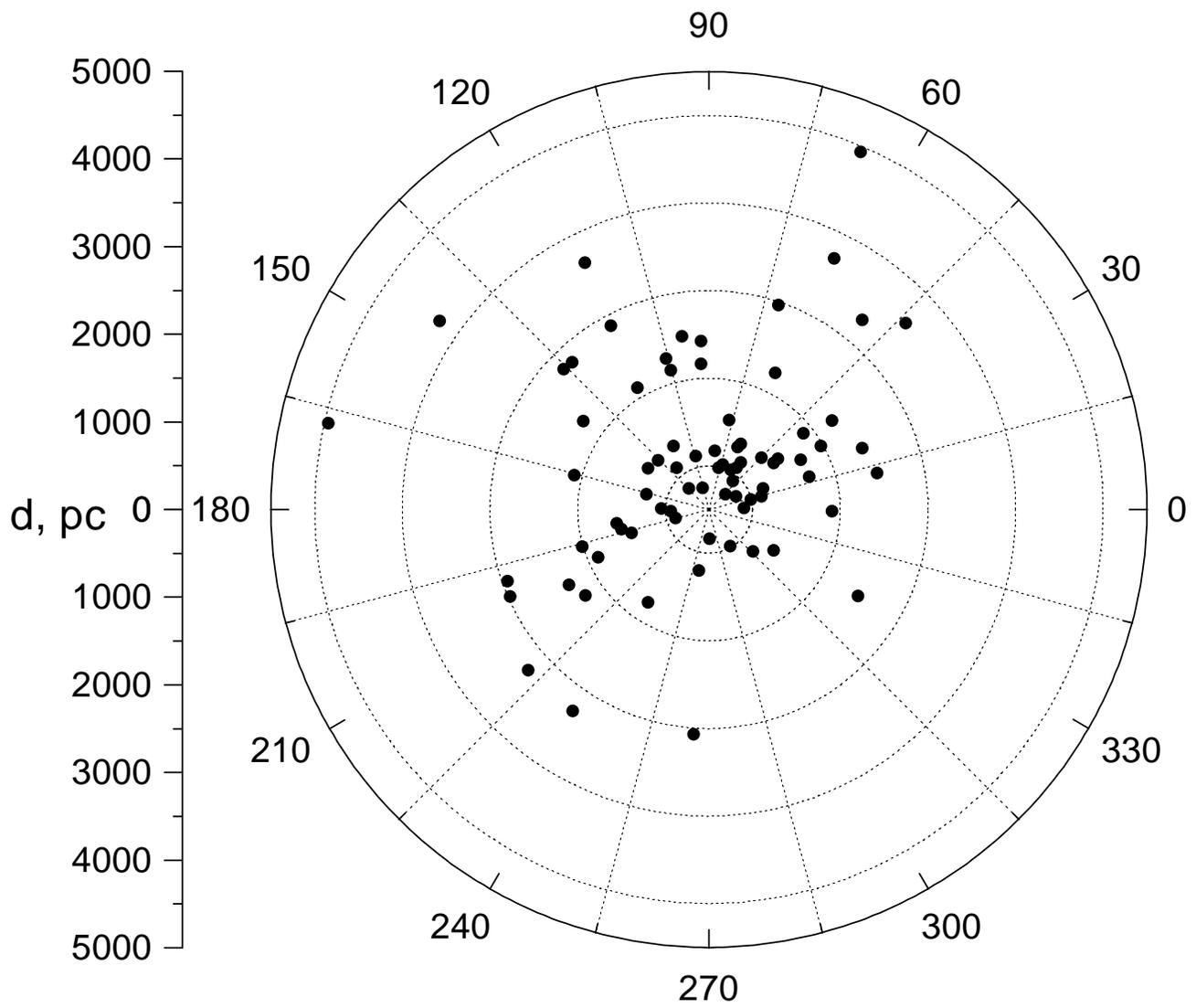}}
\caption[]{The distribution of the program Cepheids in the galactic plane}
\end{figure*}
\newpage
\begin{figure*}
\resizebox{\hsize}{!}{\includegraphics{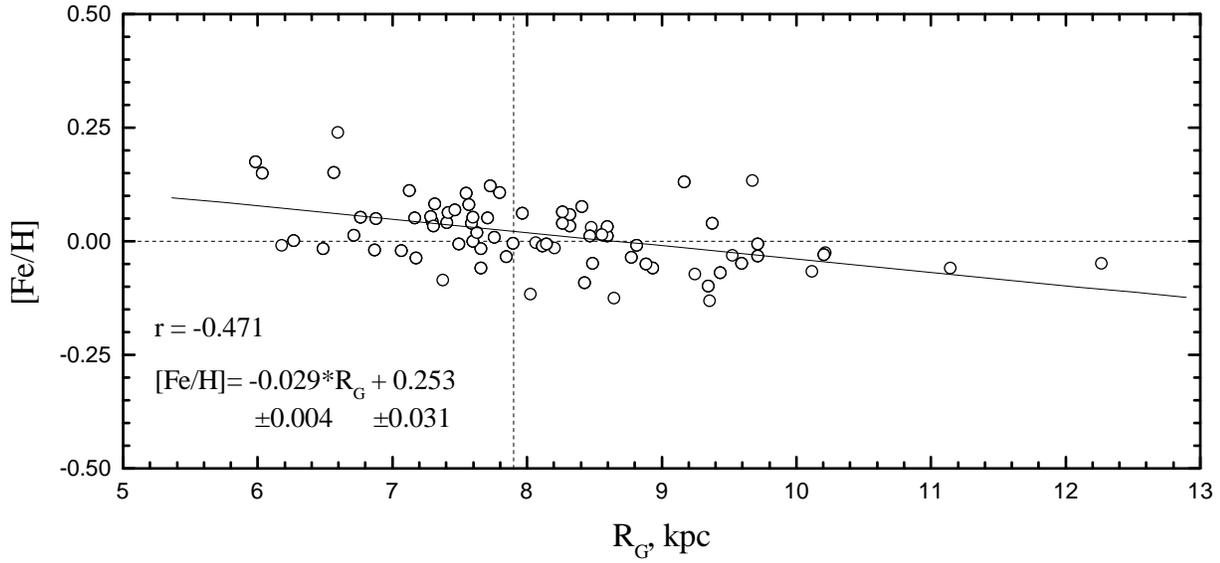}}
\caption[]{Iron abundance gradient and its linear approximation.
The position of the Sun is at the intersection of the dashed
lines}
\end{figure*}
\newpage
\begin{figure*}
\resizebox{\hsize}{!}{\includegraphics{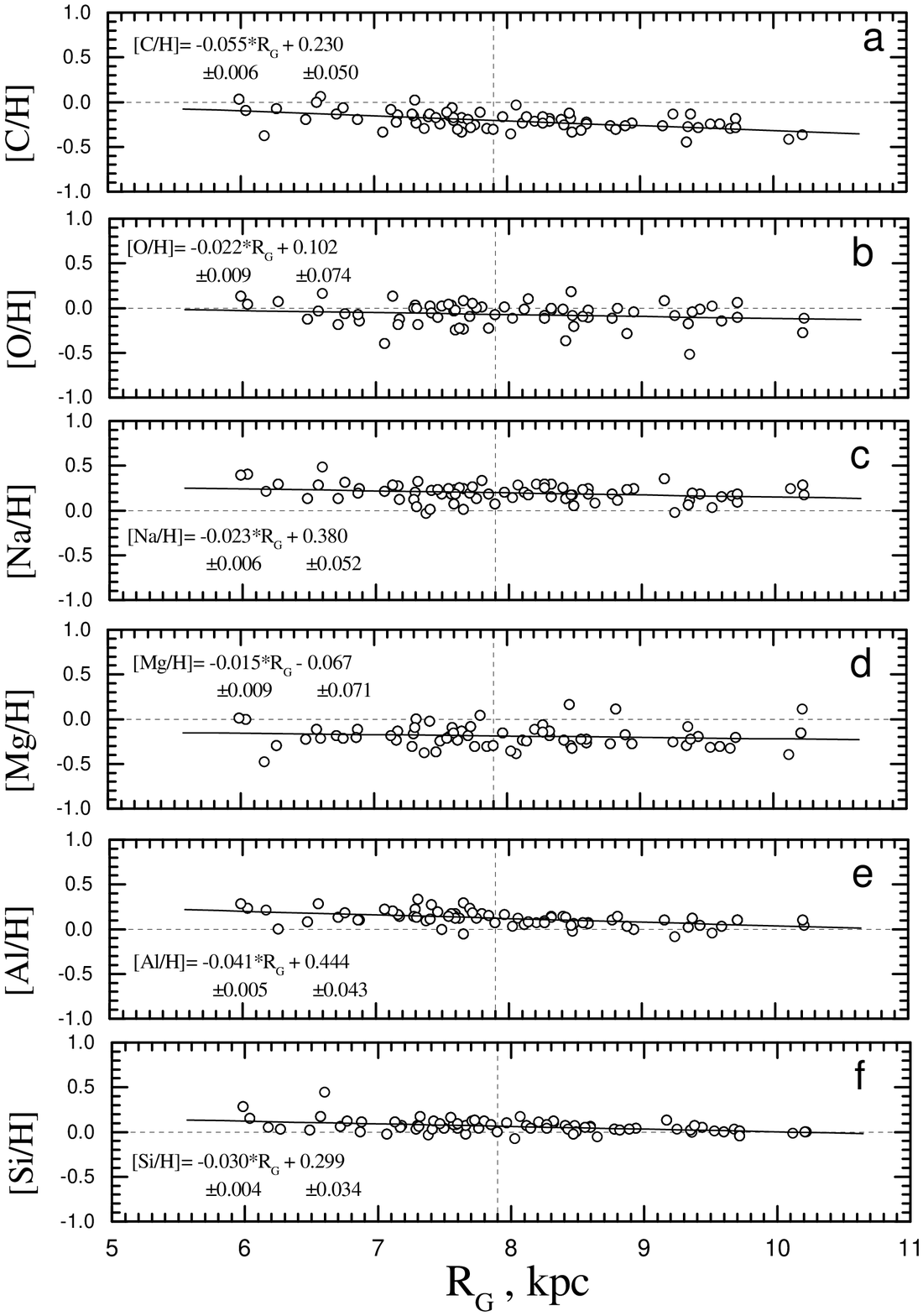}}
\caption[]{Abundance gradients for other investigated elements:
C--Si}
\end{figure*}
\newpage
\begin{figure*}
\resizebox{\hsize}{!}{\includegraphics{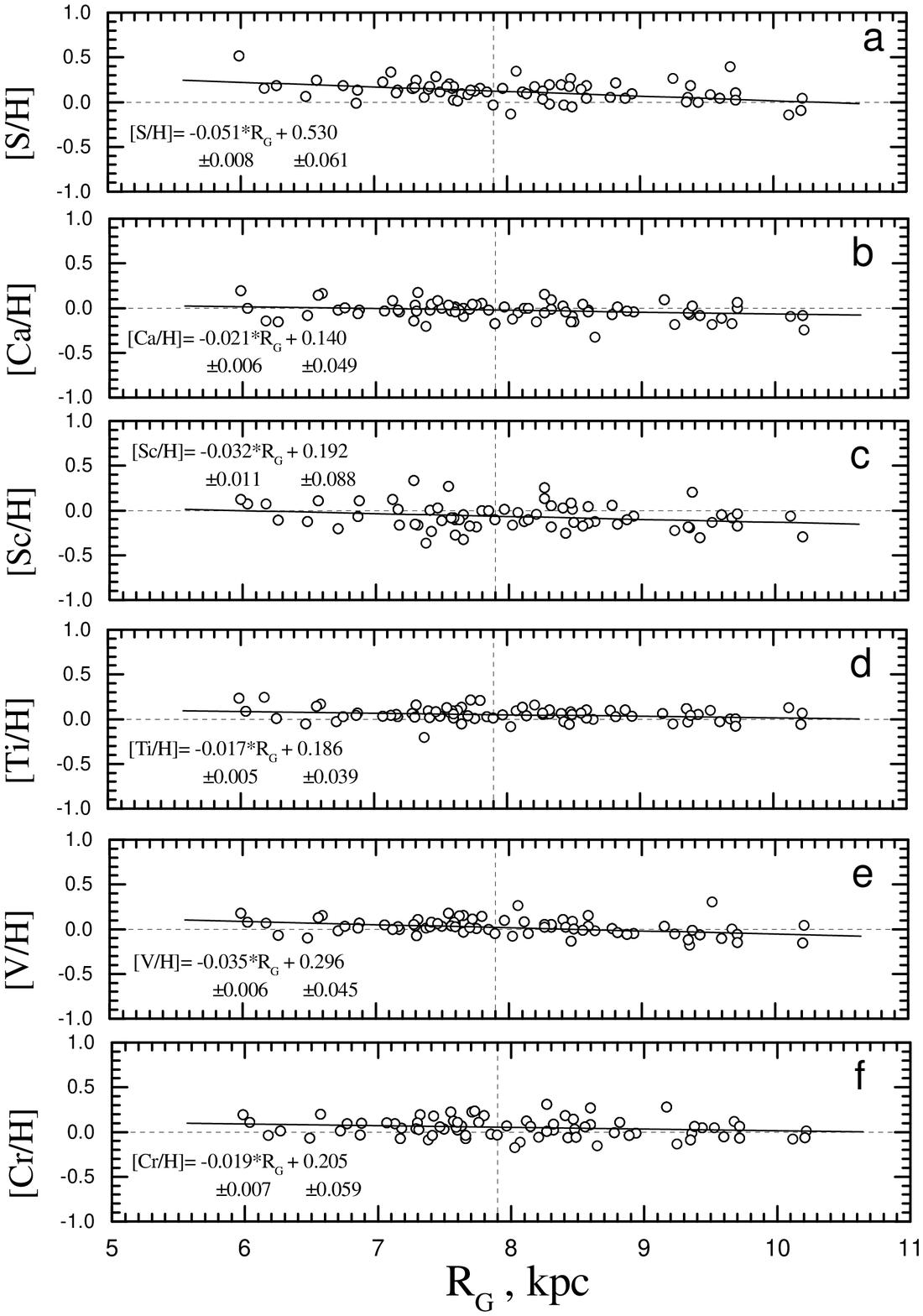}}
\caption[]{Same as Fig.6, but for S--Cr}
\end{figure*}
\newpage
\begin{figure*}
\resizebox{\hsize}{!}{\includegraphics{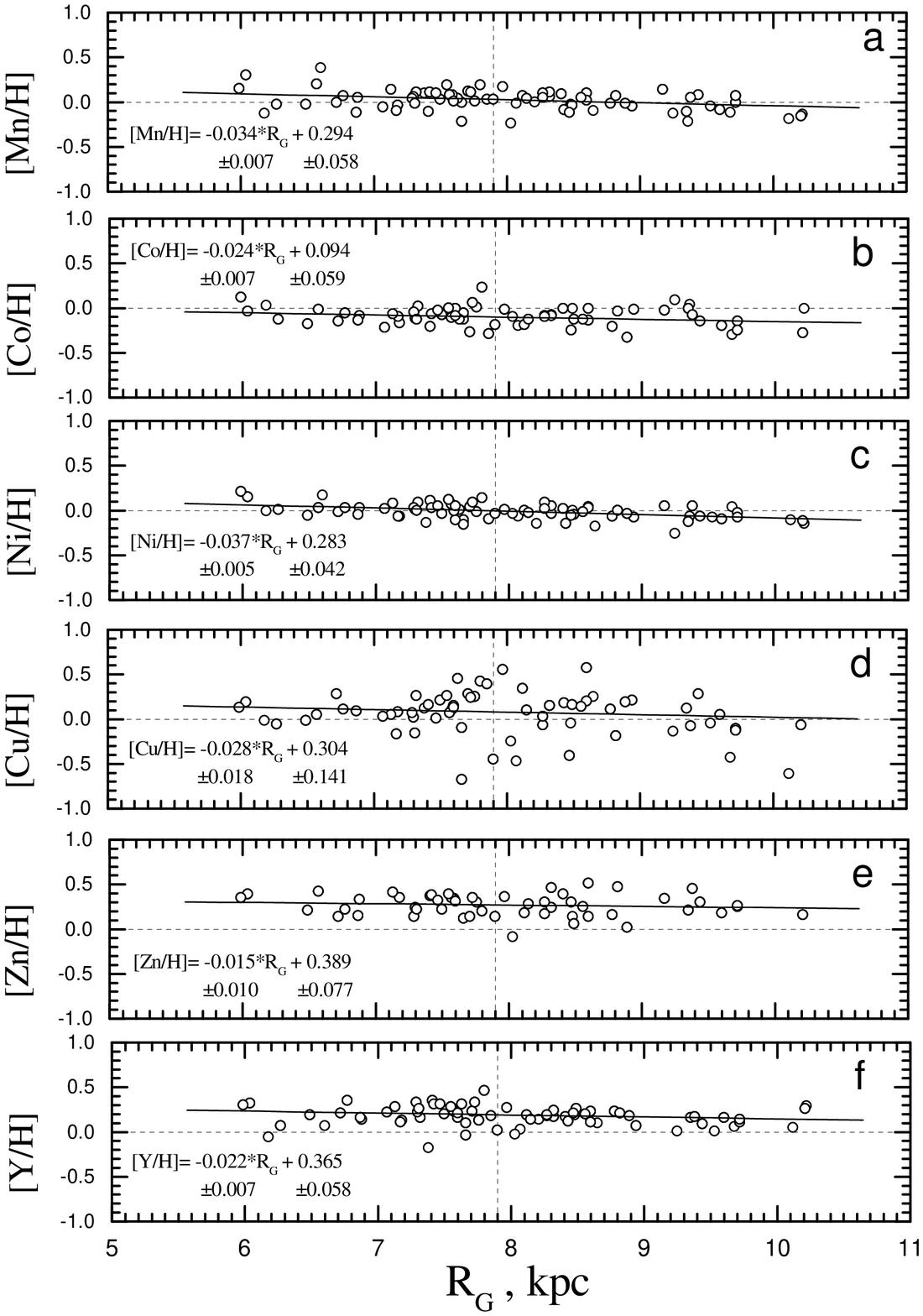}}
\caption[]{Same as Fig.6, but for Mn--Y}
\end{figure*}
\newpage
\begin{figure*}
\resizebox{\hsize}{!}{\includegraphics{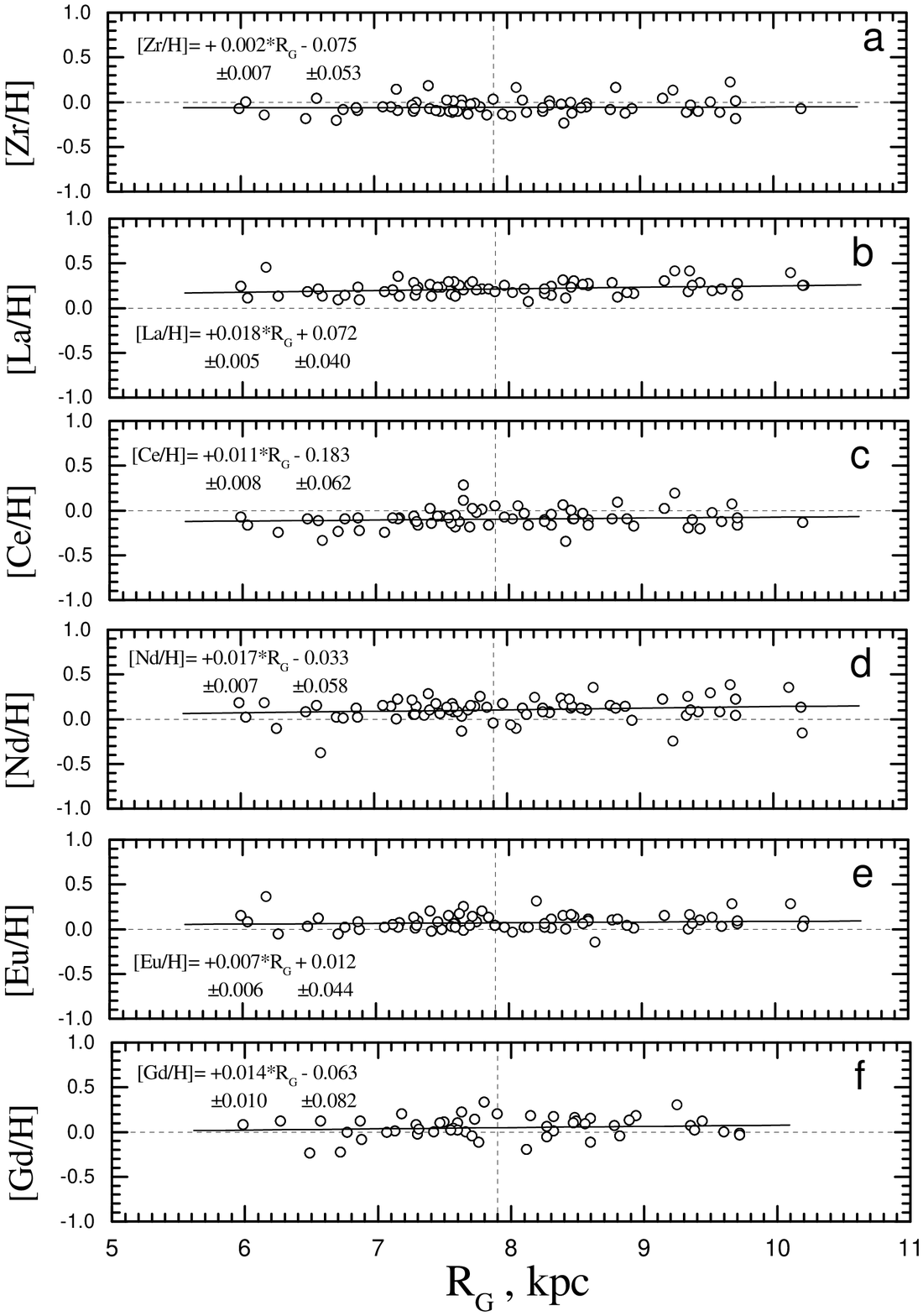}}
\caption[]{Same as Fig.6, but for Zr--Gd}
\end{figure*}
\newpage
\begin{figure*}
\resizebox{\hsize}{!}{\includegraphics{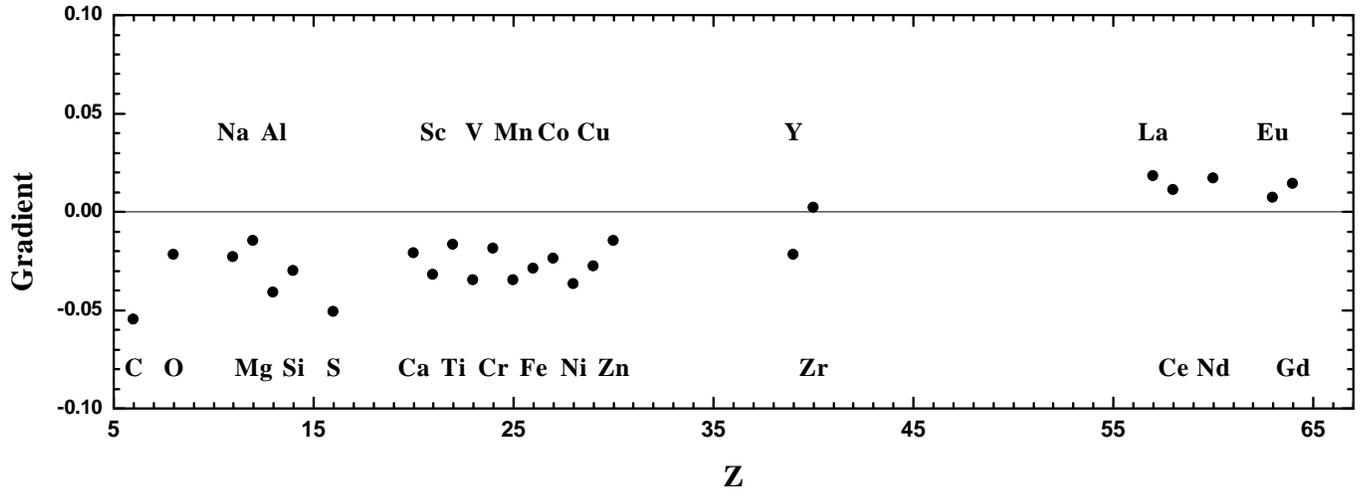}}
\caption[]{Derived gradients versus atomic number}
\end{figure*}
\newpage
\begin{figure*}
\resizebox{\hsize}{!}{\includegraphics{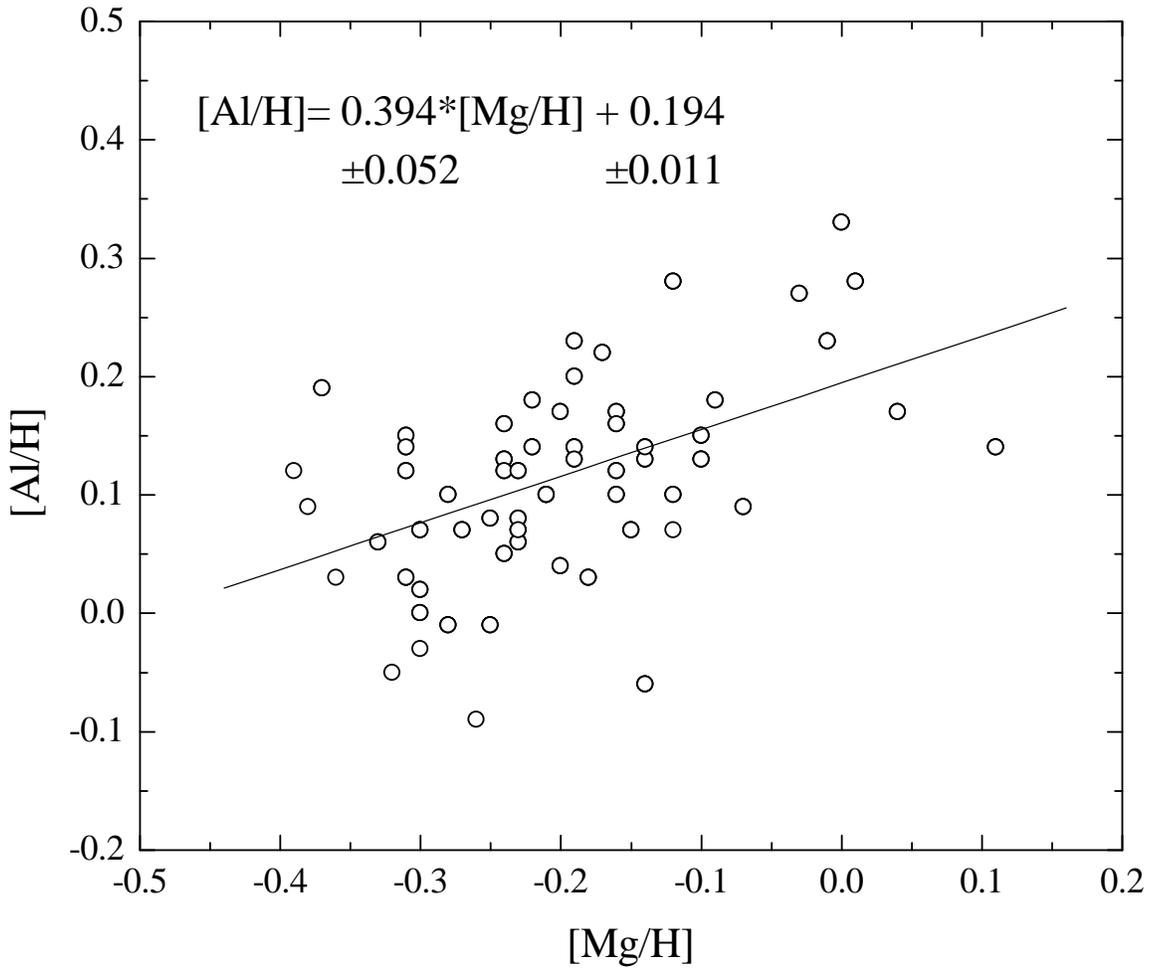}}
\caption[]{[Al/H] vs. [Mg/H] for program Cepheids}
\end{figure*}
\newpage
\begin{figure*}
\resizebox{\hsize}{!}{\includegraphics{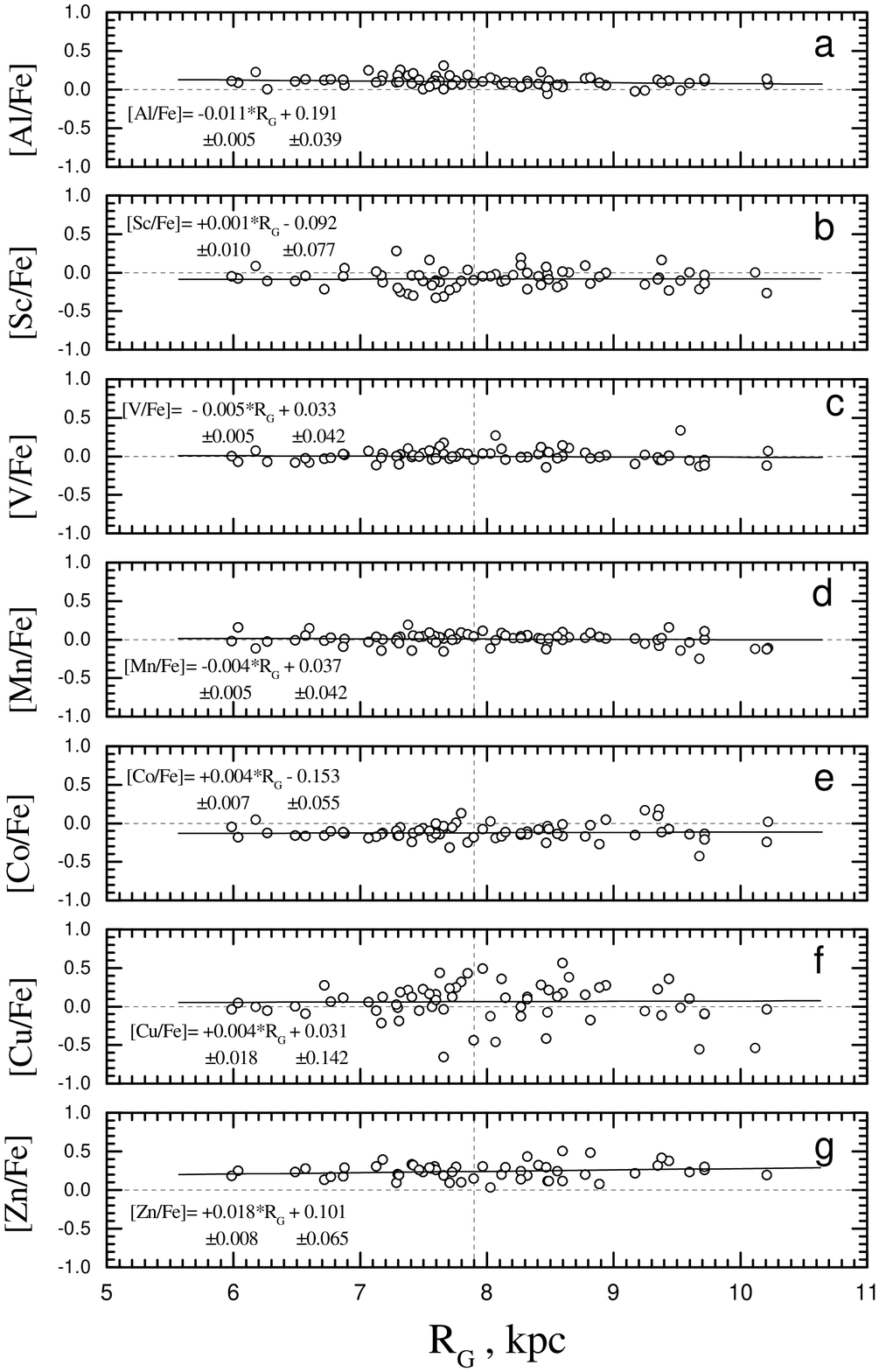}}
\caption[]{Gradients for some abundance ratios}
\end{figure*}
\newpage
\begin{figure*}
\resizebox{\hsize}{!}{\includegraphics{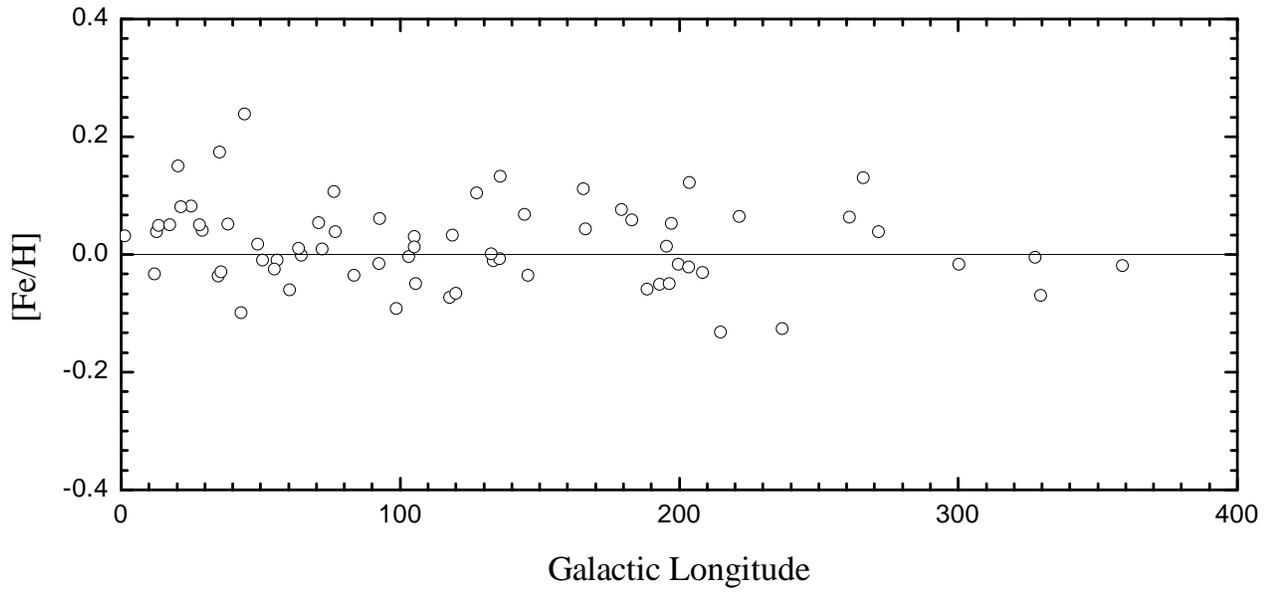}}
\caption[]{Iron abundance vs. galactic longitude}
\end{figure*}
\newpage
\begin{figure*}
\resizebox{\hsize}{!}{\includegraphics{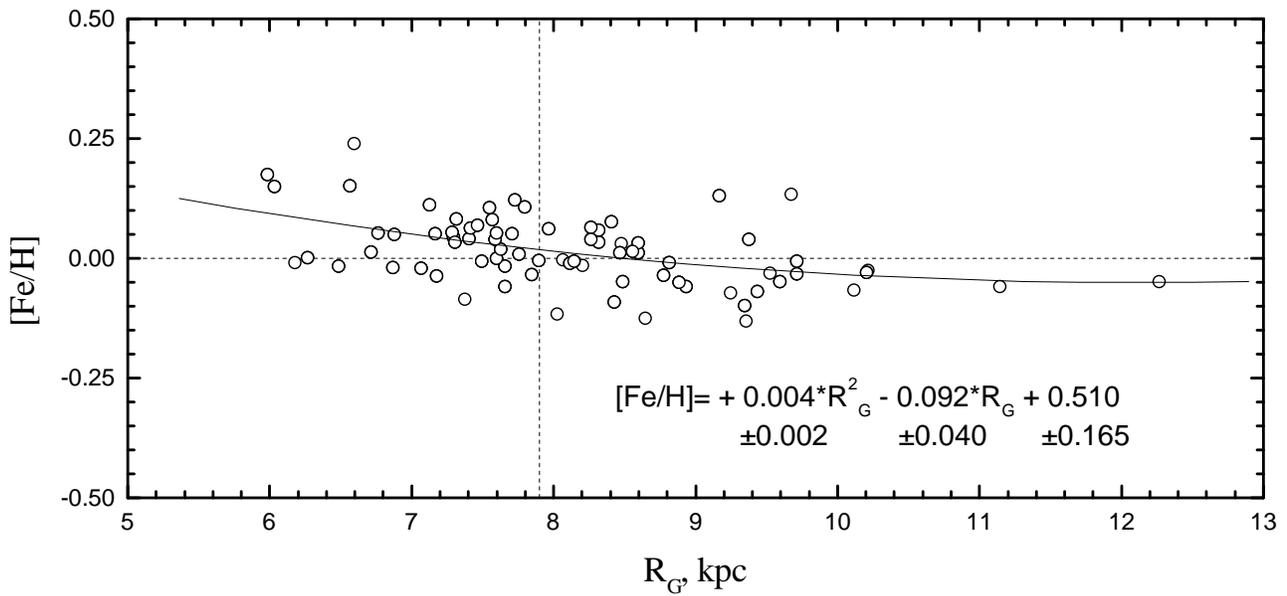}}
\caption[]{Iron abundance profile with a parabolic approximation}
\end{figure*}

\end{document}